\documentclass[journal=jpccck,manuscript=article,layout=twocolumn]{achemso}

\usepackage[version=3]{mhchem} 
\usepackage{subfig}
\usepackage{tabularx}

\usepackage{xr}
\makeatletter
\newcommand*{\addFileDependency}[1]{
  \typeout{(#1)}
  \@addtofilelist{#1}
  \IfFileExists{#1}{}{\typeout{No file #1.}}
}
\makeatother
 
\newcommand*{\myexternaldocument}[1]{%
    \externaldocument{#1}%
    \addFileDependency{#1.tex}%
    \addFileDependency{#1.aux}%
}

\myexternaldocument{supporting}

\author{Jannes Seebeck}
\affiliation[Hamburg University of Technology]{Institute of Polymers and Composites, Hamburg University of Technology, Denickestra{\ss}e 15, 21073 Hamburg, Germany}

\author{Peter Schiffels}
\affiliation[Fraunhofer IFAM]{Fraunhofer-Institut f\"ur Fertigungstechnik und Angewandte Materialforschung IFAM - Klebtechnik und Oberfl\"achen, Wiener Stra{\ss}e 12, 28359 Bremen, Germany}

\author{Sabine Schweizer}
\affiliation[Scienomics GmbH]{Scienomics GmbH, B\"urgermeister-Wegele-Stra{\ss}e 12, 86167 Augsburg, Germany}

\author{J\"org-R\"udiger Hill}
\affiliation[Materials Design]{Materials Design SARL, 42, Avenue Verdier,
92120 Montrouge, France}
\alsoaffiliation[Scienomics GmbH]{Scienomics GmbH, B\"urgermeister-Wegele-Stra{\ss}e 12, 86167 Augsburg, Germany}

\author{Robert Horst Mei\ss{}ner}
\affiliation[Hamburg University of Technology]{Institute of Polymers and Composites, Hamburg University of Technology, Denickestra{\ss}e 15, 21073 Hamburg, Germany}
\alsoaffiliation[Helmholtz-Zentrum Geesthacht]{MagIC -- Magnesium Innovation Centre, Institute of Materials Research Helmholtz-Zentrum Geesthacht, Max-Planck Str. 1, 21502 Geesthacht, Germany}
\email{robert.meissner@tuhh.de}

\title[Electrical Double Layer Capacitance of Curved Graphite Electrodes]
  {Electrical Double Layer Capacitance of Curved Graphite Electrodes}

\abbreviations{MD}
\keywords{supercapacitors, electric double layer, curved electrodes, molecular dynamics, constant potential method}

\begin{document}

\begin{tocentry}
    \vspace*{\fill}
    \includegraphics[width=0.8\textwidth]{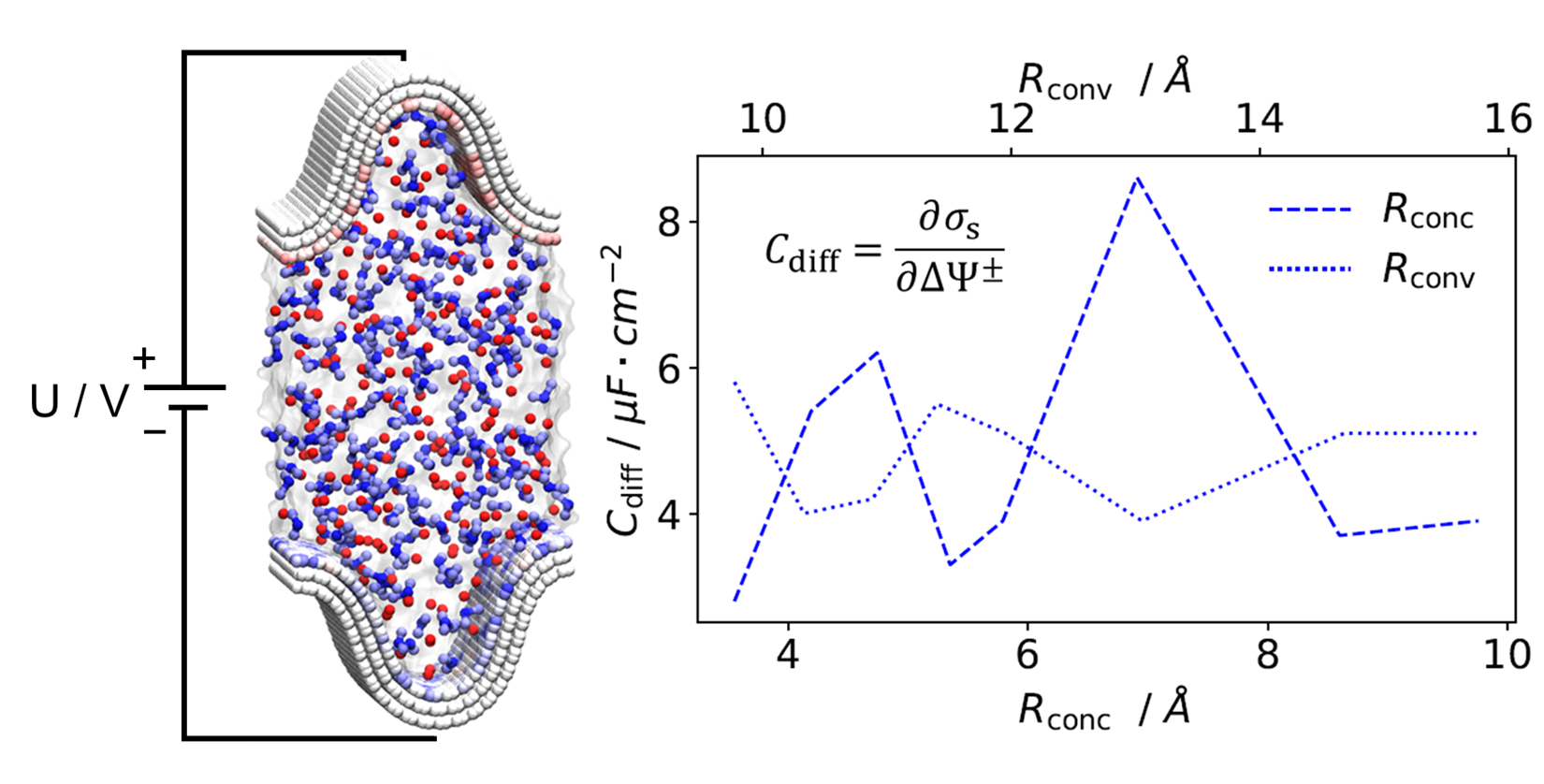}
    \vspace*{\fill}
\end{tocentry}


\twocolumn[
  \begin{@twocolumnfalse}
      This document is the unedited Author's version of a Submitted Work that was subsequently accepted for publication in Journal of Physical Chemitry C (copyright © American Chemical Society) after peer review. To access the final edited and published work see https://pubs.acs.org/doi/10.1021/acs.jpcc.9b10428. \\ 
  \end{@twocolumnfalse}]

\begin{abstract}
To improve the understanding of the relation between electrode curvature and energy storage mechanisms, a systematic investigation of the correlation between convex and concave electrode surfaces and the differential capacitance of an electrochemical double layer capacitor using molecular dynamics simulations is presented. 
Each electrode consists of three layers of curved graphene sheets with a convex and concave surface to which the constant potential method was applied.
The differential capacitance shows a fluctuating behavior with respect to the curvature radius of the convex and concave areas of the electrode. 
The reasons identified for this are differences in the geometric arrangement and solvation of the adsorbed ions as well as a steric hindrance prohibiting further charge accumulation.
Since the total differential capacitance is calculated as a weighted average of contributions from concave and convex surfaces, the influence of individual curvatures on the total capacitance is significantly reduced for the total electrode surface.
\end{abstract}

\section{Introduction}

Electrochemical double layer capacitors (EDLCs) or supercapacitors are promising energy storage devices characterized by a high power density, short charging times and a long service life. 
However, their disadvantage is a relatively low energy density compared to traditional batteries.\citep{Simon2008MaterialsCapacitors}
The energy storage mechanism in an EDLC is based on an electrostatic field formed at the interface between a conductive electrode and an electrolyte, the so-called Helmholtz double layer, with a separation of charge in the order of a few {\AA}ngstr\"oms -- several orders smaller than in a conventional capacitor.
By using organic electrolytes \citep{Wang2012,Merlet2013}, which provide a low viscosity and higher ionic conductivity as compared to ionic liquids at room temperature, EDLCs can be operated in a very large electrochemical window.\citep{Forse2016, GONZALEZ20161189, Lin2009, Merlet2013a} 
In addition to the electrolyte composition, the efficiency of the EDLC is strongly determined by the electrode material.
In particular the chemical and physical properties of carbon-based materials, such as the high specific surface area, the good electric conductivity, the high chemical stability and the wide operating temperature range, make them suitable candidates for electrodes.\citep{Wang2012,Zhang2009}
In order to develop new carbon materials with increased capacitance, it is thus crucial to understand how carbon structures affect charge storage mechanisms.\citep{Forse2016, Wang2012, GONZALEZ20161189}

Capacitances as a function of the mean pore size of several porous carbon-based structures were previously determined experimentally as well as from simulations.\citep{Largeot2008, Huang2008, Feng2011a, YoungseonShim2010,Kondrat2012,Merlet2012,doi:10.1021/jz3004624,Schweizer2019}
By limiting the electrolyte contact only to the convex part of an electrode, e.g.,\,by using the outer surfaces of fullerenes or carbon nanotubes (CNT), a reduction in capacitance was observed.\citep{doi:10.1021/jz401472c,Feng2013, Bo2018}
In all these studies, however, only the influence of either purely concave or convex geometric structures were investigated.
In complex porous structures, e.g.,\,amorphous carbons on the contrary, influences from different geometries are always intertwined.
Hence, it is practically impossible to allow rigorous conclusions about influences of edges, pores, curvatures and their combination on the differential capacitance from simulations of amorphous carbons alone.
Thus, this work attempts to investigate individual contributions of convex and concave electrode curvatures to the differential capacitance of the entire electrode and establish a link to more complex porous structures.

\section{Computational Details}

\begin{figure}
  \centering
  \includegraphics[width=0.33\textwidth]{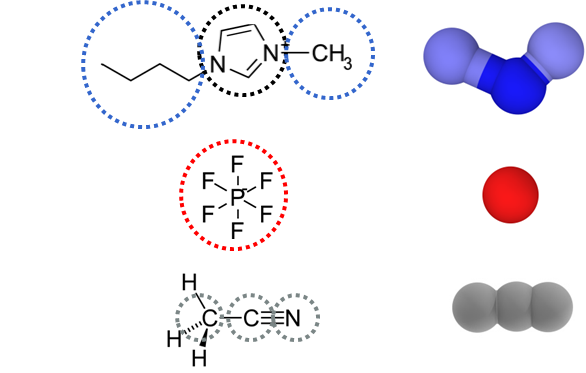}
  \caption{Coarse grained model of the electrolyte used in the simulations. Parameters are listed in Table \ref{tbl:ele_parm}. \label{fig:ele_mod}}
\end{figure}

The calculations were carried out within the framework of molecular dynamic simulations using the Large-scale Atomic/Molecular Massively Parallel Simulator (LAMMPS)\citep{lammps}.
The simulation cell consisted of two curved graphitic electrodes enclosing an electrolyte. The electrolyte was made of a 1.5\,M 1-n-butyl-3-methylimidazolium hexafluorophosphate ([BMI]$^+$-[PF$_6$]$^-$) solution in acetonitrile (ACN) and was represented by a coarse grained model as summarized by Merlet et al. \citep{Merlet2013}.
Fig.\,\ref{fig:ele_mod} shows schematically the structure of the coarse grained molecules. 
The corresponding simulation parameters are summarized in Table\,\ref{tbl:ele_parm}.

\begin{table*}[htbp]
  \centering
  \caption{Coarse grained model parameters for the molecules represented in Fig. \ref{fig:ele_mod} as reported by Merlet et al.\,\citep{Merlet2013}. \label{tbl:ele_parm}}
    \begin{tabular}{cccccccccccccccc}
    \hline
      Site & \multicolumn{2}{c}{Imi} & \multicolumn{2}{c}{Met}  & \multicolumn{2}{c}{But} & \multicolumn{2}{c}{[PF$_6$]$^-$} & \multicolumn{2}{c}{N} & \multicolumn{2}{c}{C} & \multicolumn{2}{c}{Me}\\
      \hline
      $q$ / $e$                          & \multicolumn{2}{c}{0.4374} & \multicolumn{2}{c}{0.1578} & \multicolumn{2}{c}{0.1848} & \multicolumn{2}{c}{-0.78} & \multicolumn{2}{c}{-0.398} & \multicolumn{2}{c}{0.129} & \multicolumn{2}{c}{0.269}\\
      $M$ / $\mathrm{g \cdot mol^{-1}}$            & \multicolumn{2}{c}{67.07} & \multicolumn{2}{c}{15.04} & \multicolumn{2}{c}{57.12} & \multicolumn{2}{c}{144.96} & \multicolumn{2}{c}{14.01} & \multicolumn{2}{c}{12.01} & \multicolumn{2}{c}{15.04} \\
      $\sigma_i$ / \AA                & \multicolumn{2}{c}{4.38} & \multicolumn{2}{c}{3.41} & \multicolumn{2}{c}{5.04} & \multicolumn{2}{c}{5.06} & \multicolumn{2}{c}{3.30} & \multicolumn{2}{c}{3.40} & \multicolumn{2}{c}{3.60} \\
      $\epsilon_i$ / $\mathrm{kJ \cdot mol^{-1}}$ & \multicolumn{2}{c}{2.56} & \multicolumn{2}{c}{0.36} & \multicolumn{2}{c}{1.83} & \multicolumn{2}{c}{4.71} & \multicolumn{2}{c}{0.42} & \multicolumn{2}{c}{0.42} & \multicolumn{2}{c}{1.59} \\
      \hline
    \end{tabular}
\end{table*}

For comparison of the performance, additional simulations with flat graphitic electrodes were carried out as well.

The simulation system was set up in three steps: First, Packmol\,\citep{Szalay2010} was used to obtain a statistical distribution of the coarse grained molecules in a bulk electrolyte simulation cell. 
After a 0.02\,ns equilibration run in the NVE ensemble, followed by a 0.02\,ns NVT simulation at 300\,K, the density of the bulk electrolyte was adjusted by a 2\,ns NPT simulation at 300\,K and 1.0\,bar using the Nos\'{e}-Hoover thermostat/barostat. 
In all simulations the bonds of the ACN as well as bonds and angles of the [BMI]$^+$ were constrained using the SHAKE algorithm, whilst for the bond angle of ACN a stiff spring was used to fix the angle to 180$^\circ$. 
The electrostatic interactions were calculated using the particle-particle particle-mesh method (pppm) along with a cutoff radius of 12\,\AA~for the short ranged potentials.

In the second step, a three layered planar graphitic model served as initial structure for modeling the wave-like electrodes.
In order to define the concave and convex radii of the middle layer in the final curved structure the length of the initial simulation cell perpendicular to the \textit{zig-zag} direction of the planar graphene layers should be equal to the perimeter of a corresponding carbon nanotube (CNT) with the same radius.
The wave-like structure resulted from a deformation of that cell dimension to twice the diameter of the CNT equivalent.
There is a series of carbon potentials that can be used for the simulation of carbon based electrode structures and are described by de Tomas et al.\citep{DeTomas2016}.
We used the LCBOP \citep{PhysRevB.68.024107} potential in a NVT simulation at 300\,K for 0.05\,ns and a following energy minimization. 

The final electrode radii for the middle layer, $R_\mathrm{mid}$, as well as for the concave and convex surface areas which are in contact with the electrolyte, $R_\mathrm{conc}$ and $R_\mathrm{conv}$, are presented in Tab.~\ref{tbl:radii_concave_convex}.

\begin{table}[h]
\centering
  \caption{\textnormal{Radii of the concave, $R_\mathrm{conc}$, and convex, $R_\mathrm{conv}$, parts of the inner layer in contact with the electrolyte as well as the radius of the middle layer ($R_\mathrm{mid}$). Radii are obtained by fitting circles to convex and concave areas of the electrode.\label{tbl:radii_concave_convex}}}
    \begin{tabular}{cccccccc}
    \hline
      \multicolumn{2}{c}{$R_\mathrm{conv}$ / \AA} & \multicolumn{2}{c}{$R_\mathrm{conc}$ / \AA} & \multicolumn{2}{c}{$R_\mathrm{mid}$ / \AA}\\
      \hline
      \multicolumn{2}{c}{9.77} & \multicolumn{2}{c}{3.55} & \multicolumn{2}{c}{4.07}\\
      \multicolumn{2}{c}{10.33} & \multicolumn{2}{c}{4.19} & \multicolumn{2}{c}{4.75}\\
      \multicolumn{2}{c}{10.88} & \multicolumn{2}{c}{4.74} & \multicolumn{2}{c}{5.42}\\
      \multicolumn{2}{c}{11.40} & \multicolumn{2}{c}{5.35} & \multicolumn{2}{c}{6.10}\\
      \multicolumn{2}{c}{11.95} & \multicolumn{2}{c}{5.79} & \multicolumn{2}{c}{6.78}\\
      \multicolumn{2}{c}{13.05} & \multicolumn{2}{c}{6.91} & \multicolumn{2}{c}{8.14}\\
      \multicolumn{2}{c}{14.68} & \multicolumn{2}{c}{8.60} & \multicolumn{2}{c}{10.17}\\
      \multicolumn{2}{c}{15.76} & \multicolumn{2}{c}{9.76} & \multicolumn{2}{c}{11.53}\\
      \hline
    \end{tabular}
\end{table}

\begin{figure}
  \centering
  \includegraphics[width=0.4\textwidth]{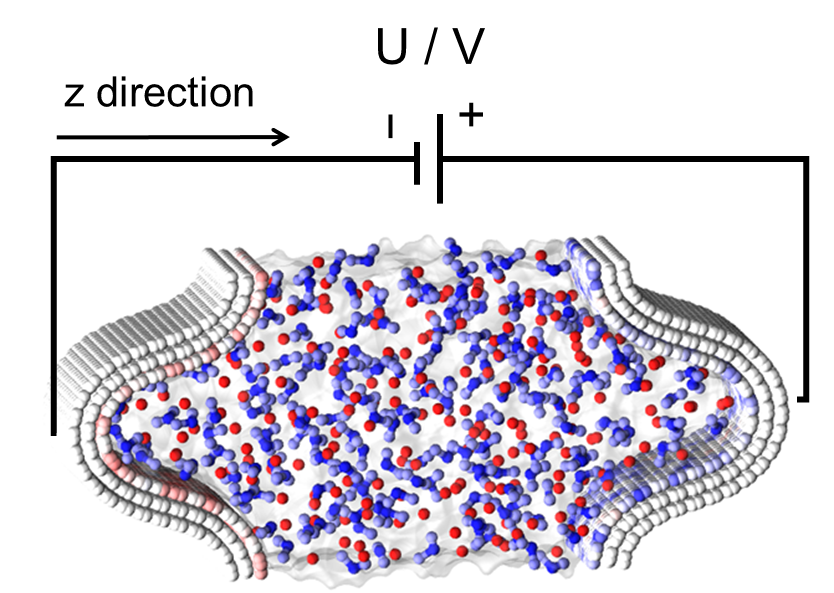}
  \caption{A model of an electrochemical cell used in the simulations. Two curved graphitic electrodes are separated by an organic electrolyte. An electrical potential $U$ is applied by employing the constant potential method. ACN is represented as a gray transparent fluid, [BMI]$^+$ and [PF$_6$]$^-$ are represented by blue and red particles, respectively.\label{fig:sim_mod}}
\end{figure}

In the third step, the electrolyte has to adapt the contour of the electrodes while a realistic bulk density in the center of the simulation box is preserved. 
Thus, the curved electrodes have to be pressed against the bulk electrolyte from both sides in order to represent experimental conditions at the given temperature and concentration according to the results of Huo et al. \citep{doi:10.1021/je700266n}.
An initial displacement of the electrodes was achieved using a NVT ensemble at 400\,K applied to the pure ionic liquid and at 300\,K for the organic electrolyte with a simulation time of 100\,ps and a time step of 2\,fs. 
After the initial compression, the displacement of the electrodes was coupled to the experimentally determined bulk density of 0.96\,g\,cm$^{-3}$ in the center of the electrolyte.
During the displacement of the electrodes the density was evaluated at each time step. The displacement was successively adjusted during the simulation in order to converge to the experimental density. 
In this case, a final error of the density of $\pm$0.03\,g\,cm$^{-3}$ was achieved. 
The final structure of an electrochemical cell with curved electrodes as used in our simulations is illustrated in Fig.\,\ref{fig:sim_mod}.

The determination of the differential capacitance was carried out in the NVT ensemble at 300\,K for the solvated liquid and at 400\,K for the pure ionic liquid.
After an equilibration run for 0.2\,ns, the data was recorded for 3.8\,ns at the respective electrode potential. 
In order to apply the defined electric potentials (0\,V, 1\,V, 2\,V, 3\,V, 4\,V and 5\,V) the constant potential (CONP) method, as it is described by Wang et al. \citep{Wang2014}, was used.
Under the assumption that the electrode material has an electrical conductivity close to that of metals, the CONP method calculates the partial charge for each electrode atom at a defined potential difference. 
In contrast to the constant charge method, the results of a CONP simulation are more realistic.\citep{Bo2018} 
Furthermore, due to the self-consistent charge determination of this method, the CONP method can be applied to arbitrary electrode geometries.\citep{Merlet2013}
A well-established approach to calculate differential capacitances for flat\citep{Merlet2011} electrodes geometries is to solve numerically the Poisson equation along the z-axis to calculate the potential drop.
Merlet et al.\citep{Merlet2012} and Lu et al.\citep{app7090939} used this approach for the potential drop calculation of supercapacitors with non-planar electrodes as well.
By using the time average of the positive or negative surface charge density $\langle \sigma^{\pm} \rangle$ of the CONP simulations as a function of the potential drop $\Delta \Psi$, the differential capacitance of the negative and positive electrode can be calculated with the equation 
\begin{equation} \label{eq:1}
    C_\mathrm{diff} = \frac {\delta \langle \sigma^{\pm} \rangle}{\delta \Delta \Psi}
\end{equation}
where the potential drop $\Delta \Psi$ is defined as the difference of the applied potential at the positive or negative electrode $\Psi^{\pm}$ and the potential in the middle of the electrolyte region $\Psi^\mathrm{bulk}$ ($\Delta \Psi = \Psi^{\pm} - \Psi^\mathrm{bulk}$). $\Psi^\mathrm{bulk}$ can be obtained by solving the Poisson equation along the direction which passes through both electrodes:
\begin{equation} \label{eq:2}
    \Psi (z) = \Psi (z_0) - \frac{1}{\epsilon_0} \int_{z_0}^z dz^\prime \int_{-\infty}^{z^\prime} dz^{\prime \prime} \rho (z^{\prime \prime}) 
\end{equation}
where $\Psi (z_0) = \Psi^{\pm}$ is the boundary condition and $\rho$ the charge density time average of an infinitesimal thin slab in $z$-direction.

\section{Results and Discussion}

\begin{figure*}
  \centering
  \subfloat[]{\includegraphics[width=0.33\textwidth]{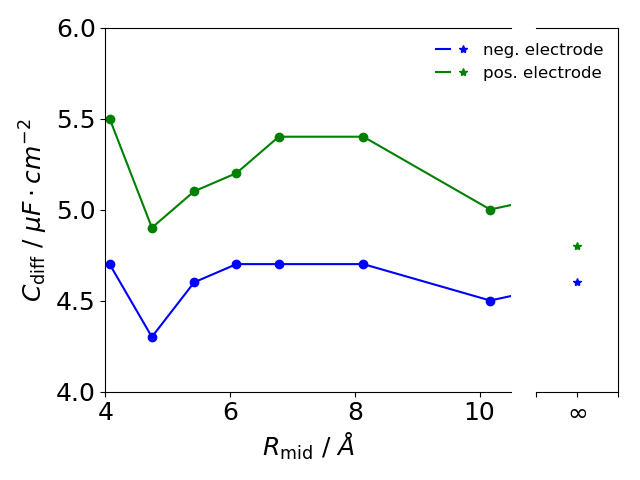}}
  \subfloat[]{\includegraphics[width=0.33\textwidth]{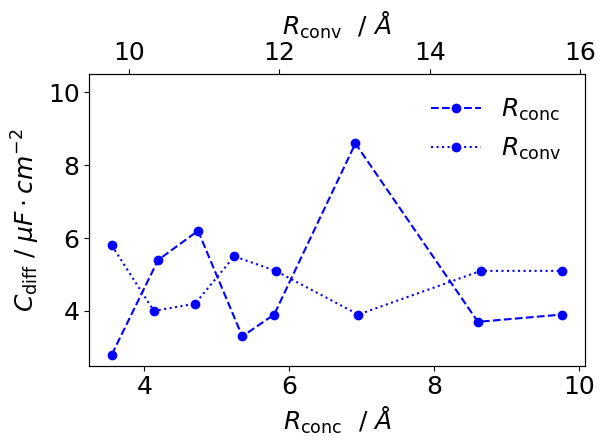}}
  \subfloat[]{\includegraphics[width=0.33\textwidth]{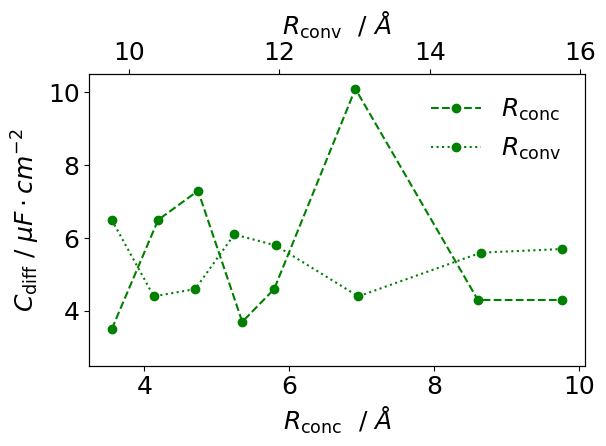}}
  \caption{(a) Differential capacitances of the positive and negative electrode as a function of the curvature radius of the middle layer, $R_\mathrm{mid}$ (the differential capacitance of flat electrodes is given by stars) and differential capacitances of concave and convex areas of negatively (b) and positively (c) charged electrodes as a function of the curvature radii, $R_\mathrm{conc}$ and $R_\mathrm{conv}$. \label{fig:diff_capa_curve_split}}
\end{figure*}

A linear regression of the time-averaged surface charges on the electrode against the potential drop allows the successive determination of the differential capacitance of each electrode (see SI for a detailed description of the approach).
Fig.\,\ref{fig:diff_capa_curve_split}(a) shows the evolution of the differential capacitance as a function of the curvature radius of the middle layer.
The curvature of the middle layer is identical in the concave and convex area.
Thus $R_\mathrm{mid}$ can be used as a reference for the capacitance of the entire electrode. 
We observed an almost constant difference in the differential capacitance for positive and negative electrodes which can be attributed to the different size of the ions.
A decrease of the capacitance from 4.7\,$\mathrm{\mu F\,cm^{-2}}$ to 4.3\,$\mathrm{\mu F\,cm^{-2}}$ for the negative electrode and from 5.5\,$\mathrm{\mu F\,cm^{-2}}$ to 4.9\,$\mathrm{\mu F\,cm^{-2}}$ for the positive electrode with an increase of the radius of curvature, $R_\mathrm{mid}$ from 4.1\,\AA~to 4.7\,\AA~is observed.
The differential capacitance increases by 0.4\,$\mathrm{\mu F\,cm^{-2}}$ for the negative and 0.5\,$\mathrm{\mu F\,cm^{-2}}$ for the positive electrode with a further increase in the curvature radius.
Compared to flat electrodes, denoted by stars in Fig.\,\ref{fig:diff_capa_curve_split}(a), positively charged curved electrodes have an overall higher capacitance (with the exception at $R_\mathrm{mid}$\,=\,4.7\,\AA).
Negatively charged curved electrodes behave worse or equivalent to the flat electrode equivalent.
Due to the fact that the electric potentials for convex and concave areas are the same, the potential drop for both is assumed to be equal.
Resulting differential capacitances for individual concave and convex regions of both electrodes using the aforementioned assumption are shown for the negative and positive electrode in Fig.\,\ref{fig:diff_capa_curve_split}(b) and Fig.\,\ref{fig:diff_capa_curve_split}(c), respectively.
Generally, higher fluctuations of the differential capacitance for individual curvatures compared to the entire electrode appear.
The concave part of the negative electrode (dashed line in Fig.\,\ref{fig:diff_capa_curve_split}(b)) shows two very clear maxima located at 4.7\,\AA~and 6.9\,\AA~ and three minima located at 3.6\,\AA, 5.4\,\AA~and 8.6\,\AA.
The convex part, on the other hand, (dotted line in Fig.\,\ref{fig:diff_capa_curve_split}(b)) shows three maxima at 9.8\,\AA, 11.4\,\AA~and 14.7\,\AA~and fluctuates with a generally smaller amplitude. 
A similar behavior is observed for the positive electrode (see Fig.\,\ref{fig:diff_capa_curve_split}(c)). 
However, a generally higher capacitance is observed for concave positive electrode parts compared to negatively charged electrodes.
Whereas for the convex parts of the positive and negative electrodes no pronounced difference is observable.
Concave and convex surface contributions to the differential capacitance of the entire electrode depend on the ratio of the respective curved area to the entire electrode surface.
By calculating the weighted arithmetic mean from the convex and concave capacitances, in which the respective surface of the concave and convex part is used as weight, the exact differential capacitance of the full electrode is indeed ultimately obtained. 
The strong signal at $R_\mathrm{conc}$\,=\,6.9\,\AA~is thus compensated in the total electrode capacitance by the larger surface area and lower capacitance of the convex region in our particular curved electrode model.

\begin{figure*}[ht]
  \centering
  \subfloat[]{\includegraphics[width=0.85\columnwidth]{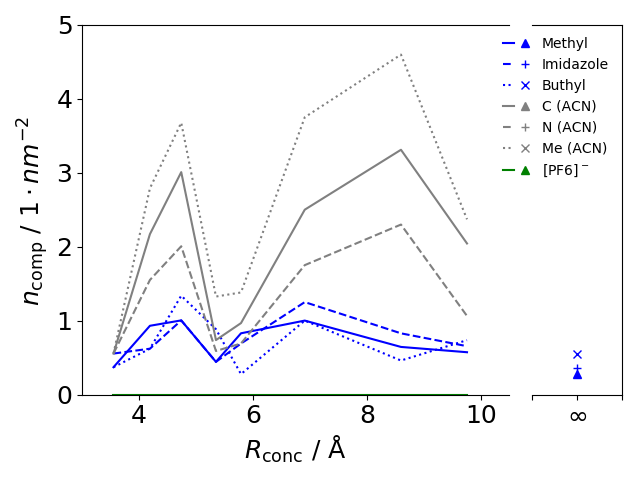}}
  \subfloat[]{\includegraphics[width=0.85\columnwidth]{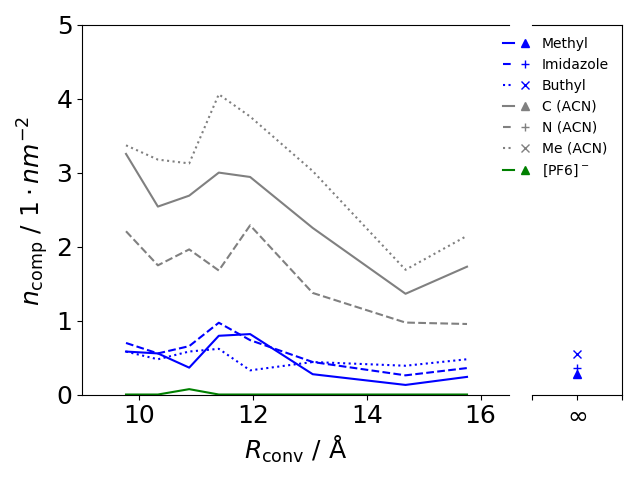}}\\
  \subfloat[]{\includegraphics[width=0.85\columnwidth]{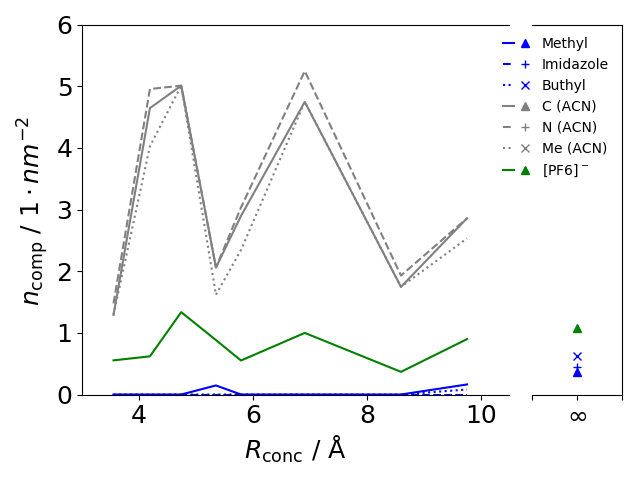}}
  \subfloat[]{\includegraphics[width=0.85\columnwidth]{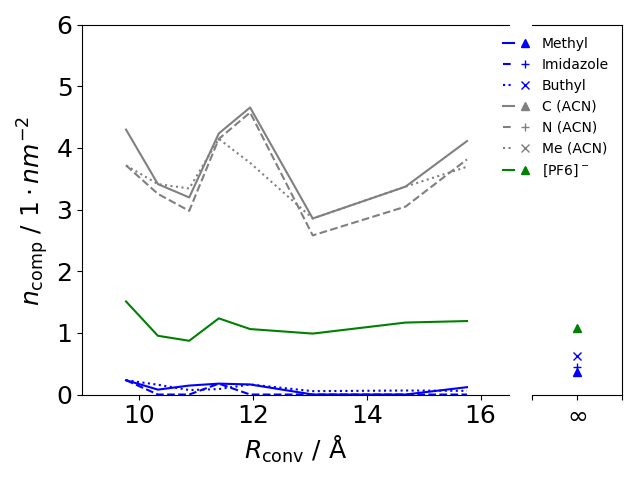}}
  \caption{Number density, $n_\mathrm{comp}$, of methyl, imidazole and buthyl groups of [BMI]$^+$, N, C and methyl groups of ACN and [PF$_6$]$^-$ in the double layer of the negatively charged concave (a) and convex (b) as well as positively charged concave (c) and convex (d) electrode areas. Results are from a simulation at 3\,V. \label{fig:number_particle_3V}}
\end{figure*}

Individual consideration of the convex and concave area allow a more rigorous interpretation of the charge storage mechanisms in these areas, while still maintaining the link to realistic graphitic structures apparent in, e.g.,\,amorphous carbons.\citep{Yang2019ElectrolyticStudy} 
At a concave radius of 3.6\,\AA, a 75\%-reduction of the solvation shell of both ion types adsorbed on the concave electrode part occurs (see Fig.\,S6) and a minimum of the differential capacitance is observed.
This behavior is similar to the decrease of the differential capacitance observed for flat electrodes when it is switched to the ionic liquid equivalent of the organic electrolyte (Fig.\,S9).
A further increase of the concave radius allows again more solvent molecules to enter the pocket (apparent in the increase of the solvation shell of the adsorbed ions, see Fig.\,\ref{fig:number_particle_3V}), effectively reducing overscreening and eventually resulting in the increase of the differential capacitance observed at 4.7\,\AA.
Multiple capacitance peaks have been also observed experimentally in \citet{Vatamanu2015} for nanoporous materials with different pore widths.
The occurrence of capacitance peaks for confined organic electrolytes was also found in the simulations of \citet{Feng2011a} using a similar organic electrolyte.
Following their argumentation, fluctuations of the differential capacitance of concave pocket-shaped electrodes could be explained as follows: 
When increasing the curvature radius, pocket sizes occur that have a larger concave surface area, but do not offer additional space for adsorbing new ions. 
In this case, the surface area is increased but no additional surface charges are induced (or they are even reduced due to the geometric shape of the electrode), leading to a capacitance minimum. 
Increasing the curvature radius further, additional ions can adsorb at the surface and another capacitance maximum is observed.
However, this argumentation is not able to explain some of the features apparent in Fig.\,\ref{fig:diff_capa_curve_split}(b) and (c), e.g.,\,the height and the difference of the second peaks observed at a concave radius of 6.7\,\AA~for negative and positive electrodes.
Thus, a more sophisticated investigation of the ionic adsorption structure and the interplay with the solvent is required to address, e.g.,\,the influence of the change of the relative permittivity due to adsorbed ACN between the electrode and the ions on the capacitance.
The number of ACN, [PF$_6$]$^-$ and [BMI]$^+$ components in the double layer at the positive and negative electrode are considered in more detail in the following to explain the capacitance peaks.
The extend of the double layer at the positive and negative electrodes were estimated from the first RDF peak of electrode carbon atoms to imidazole groups (for the negative electrode) and [PF$_6$]$^-$ (for the positive electrode). 
This method was used to determine a double layer thickness of 4.5\,\AA~for the positive and 5.0\,\AA~for the negative electrode of both flat and curved graphitic structures.
As an example, Fig.\,\ref{fig:number_particle_3V} shows the number of ACN and [BMI]$^+$ components as well as [PF$_6$]$^-$ ions within the Helmholtz layer of the concave and convex areas of the negative (a and b) and positive electrode (c and d) at 3\,V, with respect to the radius of curvature.
Results for other electrode potentials are found in the SI.
In general, number densities of ACN components show more pronounced peaks for both electrodes than for the ionic components. 
A correlation between ACN and ionic component peaks in Fig.\,\ref{fig:number_particle_3V}(a) and (c) with the concave capacitance in Fig.\,\ref{fig:diff_capa_curve_split}(b) and (c) is observed.  
However, the decrease of the ACN number at the negative convex area in Fig.\,\ref{fig:number_particle_3V}(b) at higher curvature radii shows not the same behavior as the convex capacitance in Fig.\,\ref{fig:diff_capa_curve_split}(b) with increasing radii. 
The dependence of the number density on the curvature becomes more pronounced at higher surface charge densities induced by the applied potential (see Fig.\,S10-S13).
The fluctuations of the number densities of ACN components as well as the changing ratio of [BMI]$^+$ components in the double layer of the electrodes with increasing curvature indicate a possible rearrangement and different adsorption geometries of the ions depending on the curvature of the electrode.
Due to the location of the [BMI]$^+$ charge centre between two organic side chains acting as spacers, some adsorption orientations of the [BMI]$^+$ eventually induce a higher charge on the curved electrode surface than others.
The change of the adsorption geometry and an associated structural phase transition of the ions has been also previously observed for flat electrodes at different electrode potentials.\citep{Merlet2014TheOwn}

\begin{figure*}[htb]
  \begin{tabular}{c c}
  \subfloat[]{\includegraphics[width=0.35\columnwidth]{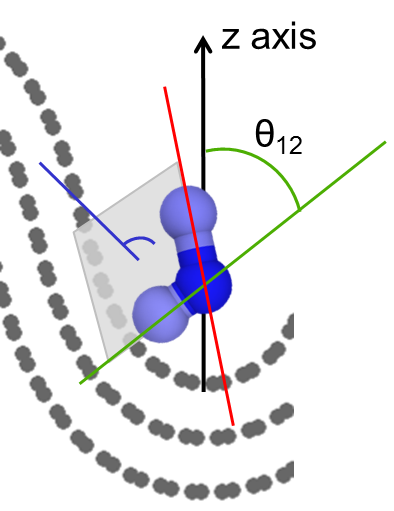}} &
  \subfloat[]{\includegraphics[width=0.85\columnwidth]{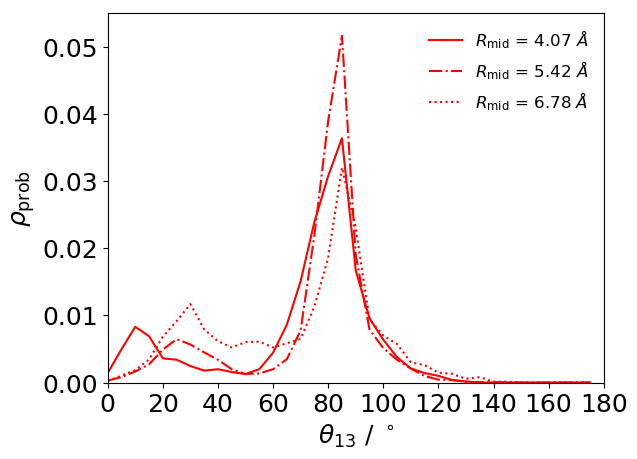}} \\
  \subfloat[]{\includegraphics[width=0.85\columnwidth]{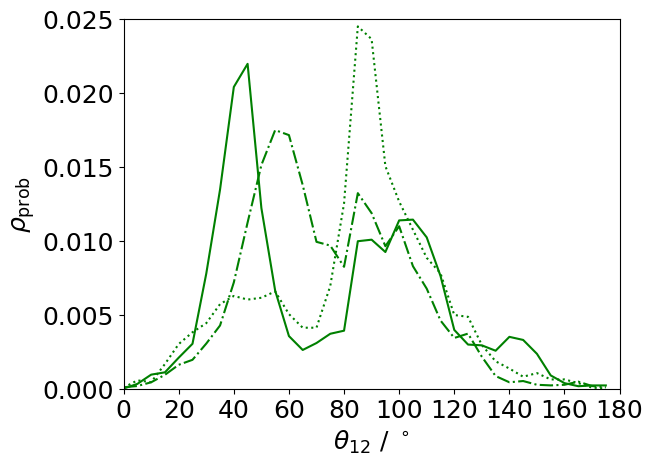}} &
  \subfloat[]{\includegraphics[width=0.85\columnwidth]{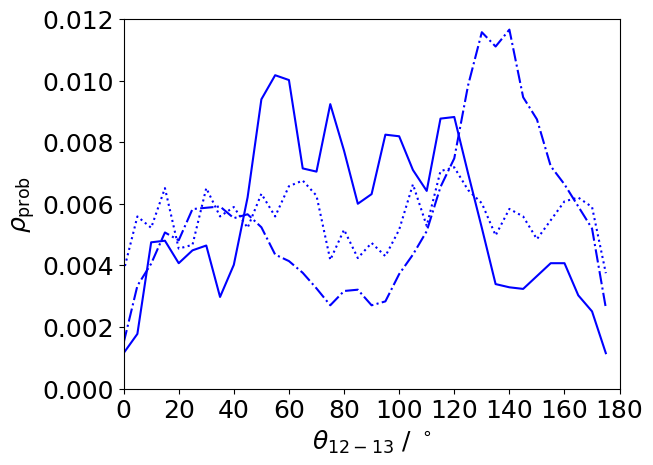}}
  \end{tabular}
  \caption{Histograms of the probability density,$\rho_{\mathrm{prob}}$, of the orientation angle between the z axis and a) the line between the imidazole and the butyl group ($\theta_{12}$) b) the line between the imidazole and the methyl group ($\theta_{13}$) and c) the normal of the plane in which all three groups of the coarsed grained [BMI]$^+$ in the double layer of the negative electrode ($\theta_{12-13}$) are located. Results are from a simulation at 3\,V. The size of the double layer was derived from RDFs between the surface carbon atoms and the imidazole groups. \label{fig:hist_orientation_angle_3V}}
\end{figure*}

A more rigorous analysis of the orientation of [BMI]$^+$ to the electrode surface is possible by defining three angles, as indicated in Fig.\,\ref{fig:hist_orientation_angle_3V}a: 
i) The angle between the line which passes through both the imidazole group and the methyl group and the z-axis of the simulation box (green line in Fig.\,\ref{fig:hist_orientation_angle_3V}a). 
ii) The angle between the line which passes through both the imidazole group and the butyl group and the z-axis of the simulation box (red line in Fig.\,\ref{fig:hist_orientation_angle_3V}a). 
iii) The angle between the normal of the plane in which the three components of [BMI]$^+$ are located and the z-axis (blue line in Fig.\,\ref{fig:hist_orientation_angle_3V}a).
Only [BMI]$^+$, whose imidazole groups are in the double layer of the negative electrode, are regarded in this calculations. 
Fig.\,\ref{fig:hist_orientation_angle_3V} shows exemplary the probability density histograms of the angles calculated from the production run for three curvatures (for histograms of other curvatures see Fig.\,S15). 

Different preferred angles for [BMI]$^+$ indicate a correlation between the adsorption orientation and the curvature radius.
Distinguishing between concave and convex area (Fig.\,S16 and Fig.\,S17) illustrates that the dependence is significantly influenced by the concave area.
The stronger dependence from concave areas could also be an explanation for the more pronounced fluctuations of the concave differential capacitance.
Thus, the [BMI]$^+$ adsorption orientations can be correlated with the number of [BMI]$^+$ molecules adsorbing on the electrode surface, explaining to some extent a capacitance increase by crowding.\citep{Bazant2011DoubleCrowding}
Some orientations favors a higher number of imidazole groups in the double layer, leading to a higher induced charge on the electrode and increased differential capacitance.
However, in order to analyze this more accurately, a more sophisticated analysis of the double layer structure at the electrode surface is necessary.
For this purpose, free energy changes associated with a change of the ion adsorption motif with varying electrode potential and curvature need to be taken into account.
Analyzing the structure of the ions and solvent at different curvatures and electrode potentials using advanced sampling techniques, e.g.,\,umbrella sampling and a weighted histogram analysis, would enable further insights into the mechanisms of the energy storage apparent in realistic carbon electrodes.
However, this goes along with an immense computational effort and would require flexible electrodes which are not (yet) possible using the CONP method in LAMMPS.

\section{Conclusion}

In this work, the influence of convex and concave electrode geometries, that are typically present in electrode materials of modern supercapacitors, on the differential capacitance has been investigated.
For this purpose, MD simulations employing a constant potential method were performed on models of electrochemical cells with curved graphitic electrodes and a commonly used organic electrolyte between them.
It was observed that the total electrode capacitance fluctuations origin from concave and convex areas of the electrode and their individual differences in their capacitance amplitudes.
Furthermore, the total differential capacitance of the entire electrode surface is calculated as the weighted average of both electrode geometries and consequently the impact of the concave area on the total differential capacitance is greater than that of the concave area. 
For this reason, differential capacitance fluctuations of the total electrode is generally attenuated.
However, the higher capacitance of the positive electrode is due to the larger influence of the concave part compared to the negative electrode.

In general, a correlation of the differential capacitance with the total number of ions and solvent in the Helmholtz layer of the electrodes was observed. 
In more detail, capacitance fluctuations are furthermore explained by a geometric reorientation of the [BMI]$^+$ and to some extent an increase or decrease of the solvation shell.
This work aims to improve the performance of supercapacitors by providing a fundamental understanding of the energy storage mechanisms in curved graphitic structures.
Thus, by favoring certain electrode geometries in the production process of amorphous carbons, cf. the structure of negatively curved schwarzites\citep{Braun2018GeneratingZeolite-templating}, it is expected to increase the capacitance of supercapacitors with this new generation electrode materials.
In this context, design criteria for the structure of electrode surfaces, e.g.,\,a favorable mean curvature radius for a specific organic electrolyte, can be developed.
%

\begin{acknowledgement}

Funded by the Deutsche Forschungsgemeinschaft (DFG, German Research Foundation) -- Projektnummer 192346071 -- SFB 986 and -- Projektnummer 390794421 -- GRK 2462. Furthermore, the authors gratefully acknowledge financial support by the German Ministry of Education and Research in the AktivCAPs project (grant no. 03SF0430B).

\end{acknowledgement}

\begin{suppinfo}

The supporting information provides details about the solvation shell calculation as well as further graphic representations for comparison with the presented data.

\end{suppinfo}

\bibliography{curve_paper}

\end{document}


This document is the unedited Author's version of a Submitted Work that was subsequently accepted for publication in Journal of Physical Chemitry C (copyright © American Chemical Society) after peer review. To access the final edited and published work see https://pubs.acs.org/doi/10.1021/acs.jpcc.9b10428. \\

\subsection{Determination of the solvation shell in the double layer}

\begin{figure}
  \centering
  \subfloat[]{\includegraphics[width=0.5\columnwidth]{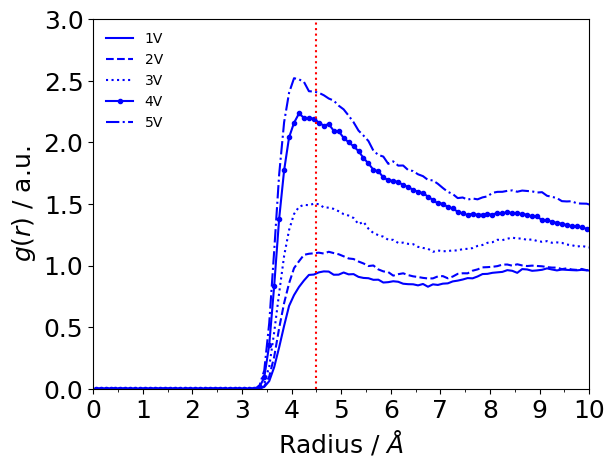}}
  \subfloat[]{\includegraphics[width=0.5\columnwidth]{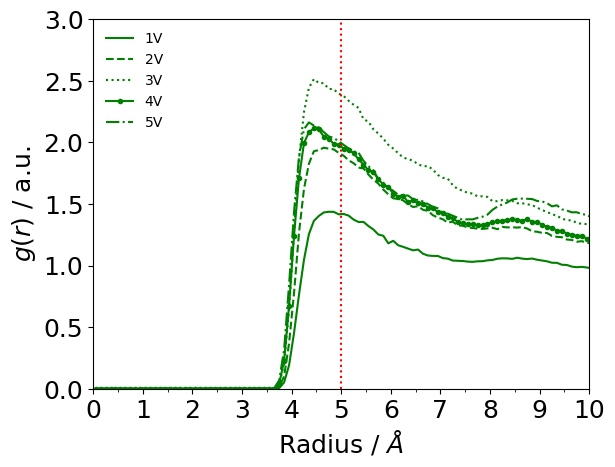}}
  \caption{Comparison of  RDFs of the carbon atoms of the flat electrode surface of the a) negative electrode with respect to the [BMI]$^+$ and b) of the positive electrode with respect to the [PF$_6$]$^-$ in the double layer of corresponding electrodes. The RDFs were taken from simulations with different electric potentials. The red dashed line shows the radius of the first solvation shell.} \label{S:fig:vergleich_rdf_flat_ele_ion}
\end{figure}

\begin{figure}
  \centering
  \subfloat[]{\includegraphics[width=0.5\columnwidth]{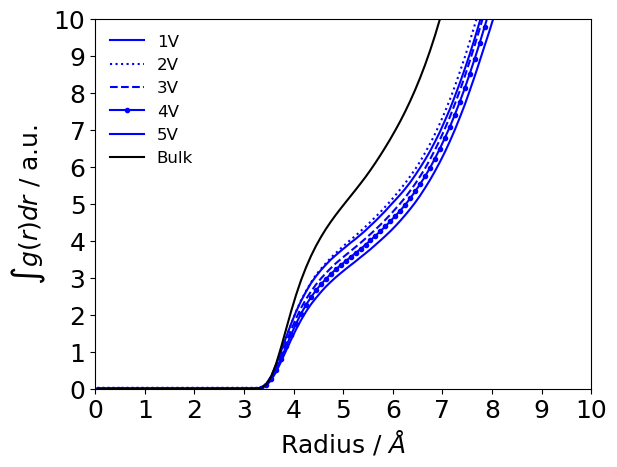}}
  \subfloat[]{\includegraphics[width=0.5\columnwidth]{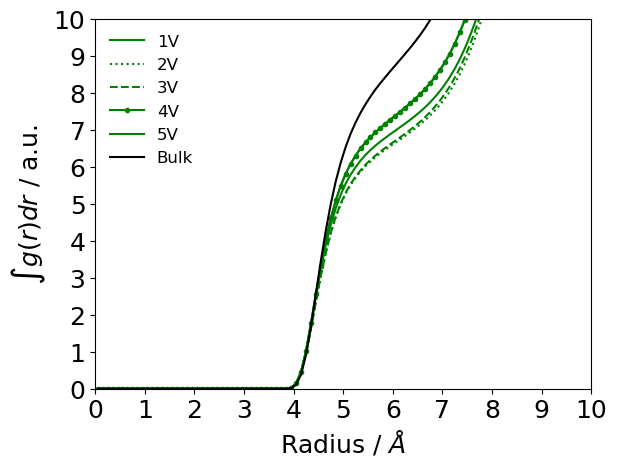}}
  \caption{Comparison of the RDF-integrals a) of the [BMI]$^+$ and b) [PF$_6$]$^-$ with respect to ACN in the double layer of flat electrodes. The RDFs were taken from simulations with different electric potentials.} \label{S:fig:coord_flat_ele}
\end{figure}

\begin{figure}
  \centering
  \subfloat[]{\includegraphics[width=0.5\columnwidth]{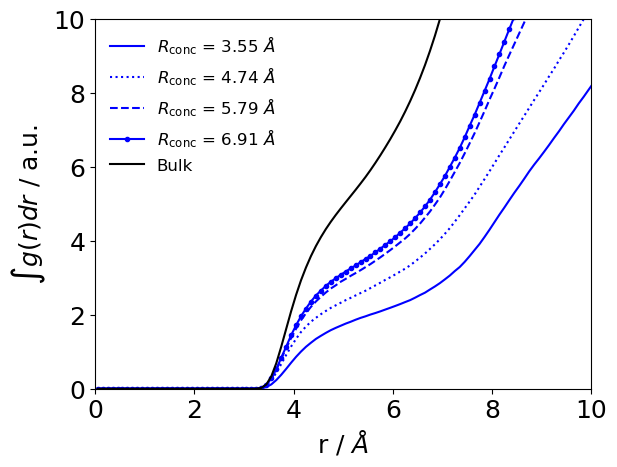}}
  \subfloat[]{\includegraphics[width=0.5\columnwidth]{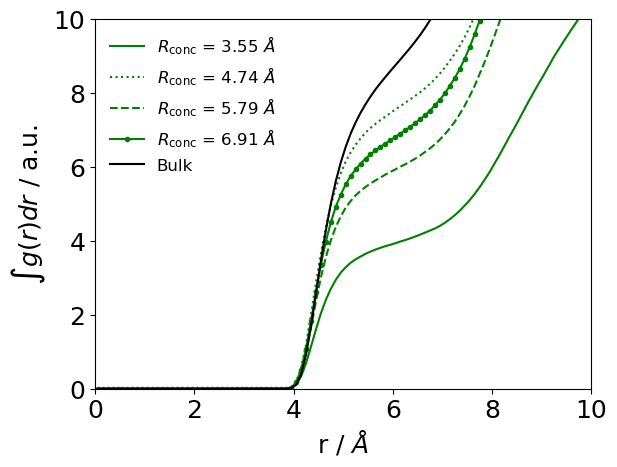}}\\
  \subfloat[]{\includegraphics[width=0.5\columnwidth]{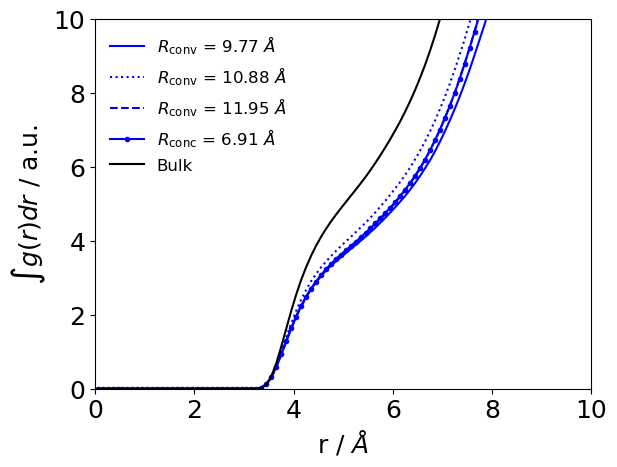}}
  \subfloat[]{\includegraphics[width=0.5\columnwidth]{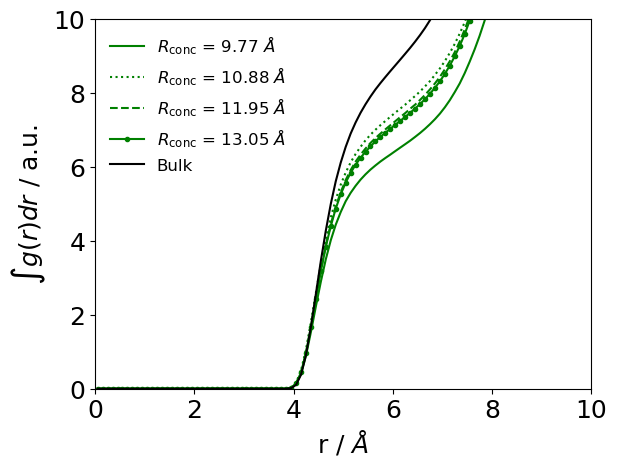}}
  \caption{Integrals of the radial distribution function of [BMI]$^+$-ACN (a,b) and [PF$_6$]$^-$-ACN (c,d) in bulk solution and in the double layer of concave (a,b) and convex (c,d) parts of the electrode with different curvatures, $R_\mathrm{conc}$ and $R_\mathrm{conv}$, respectively. [BMI]$^+$ and [PF$_6$]$^-$ results are obtained from the negative and the positive electrode, respectively. RDFs were obtained from a CONP simulation at 3\,V.} \label{S:fig:solvatshell_curve_3V}
\end{figure}

The size of the double layer was determined from the RDF's of the carbon atoms of the flat electrode surface in relation to the corresponding ions ([BMI]$^+$ for the negative electrode and [PF$_6$]$^+$ for the positive electrode). 
The radius of the first solvation shell (denoted by the red lines in Fig.\,\ref{S:fig:vergleich_rdf_flat_ele_ion}) indicates the size of the double layer.
The calculation of the RDF-integrals takes into account the imidazole group of all [BMI]$^+$ ions, in the case of the negative electrode, respectively all [PF$_6$]$^-$ coarse grained particles for the positive electrode that are within the corresponding double layer.
With regard to the ACN molecules for the determination of the solvation shell of the ions within the double layer, the nitrogen particles of the coarse grained ACN molecules were selected for the [BMI]$^+$ and the methyl groups for the [PF$_6$]$^-$. 
Using this method it is possible to determine the loss of ACN molecules of the [BMI]$^+$ and [PF$_6$]$^-$ ions when they accumulate in the double layer of the negative or positive electrode, respectively. 
Fig.\,\ref{S:fig:coord_flat_ele} illustrates the loss of the solvation shell for the ions.
A small dependence from the applied potentials to the flat electrodes and a slightly reduced number of ACN molecules are found for the first solvation shell of ACN to [BMI]$^+$ and [PF$_6$]$^-$ in the double layer as compared to the bulk organic electrolyte (solid black lines in Fig.\,\ref{S:fig:coord_flat_ele}).
The determination of the double layer size in the convex and concave areas was carried out in the same way, as it was explained for the flat electrodes, via the RDFs of the surface atoms and the imidazole particles for the negative electrode and the [PF$_6$]$^-$ particles for the positive electrode, respectively.  
Using this double layer size (4.5\,\AA~for the negative curved electrode and 5.0\,\AA~for the positive curved electrode) the number of ACN molecules in the solvation shell of the [BMI]$^+$ and [PF$_6$]$^-$ in the solvation shell were calculated with the same method which was used for the flat electrodes.
Fig.\,\ref{S:fig:solvatshell_curve_3V} shows the calculated number of ACN molecules within the first solvation shell for [BMI]$^+$ and [PF$_6$]$^-$ adsorbed in the concave and on the convex part of the electrode at 3\,V with respect to the curvature radius $R_\mathrm{conc}$ and $R_\mathrm{conv}$.
For comparison, bulk organic electrolyte integrals of the RDFs are given additionally in Fig.\,\ref{S:fig:ele_rdf} and the RDF-integrals for other potentials are given below.

\begin{figure}
  \centering
  \subfloat[]{\includegraphics[width=0.45\columnwidth]{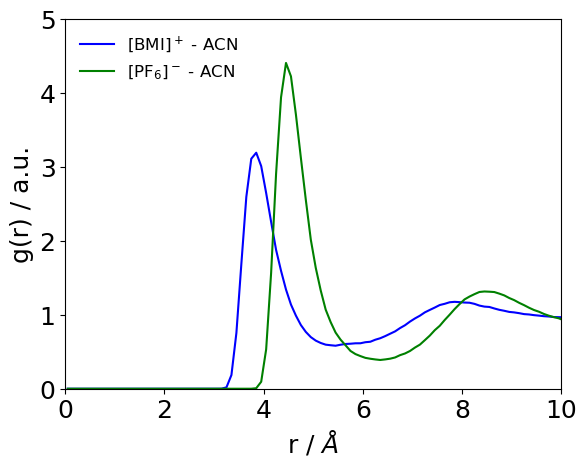}}
  \subfloat[]{\includegraphics[width=0.45\columnwidth]{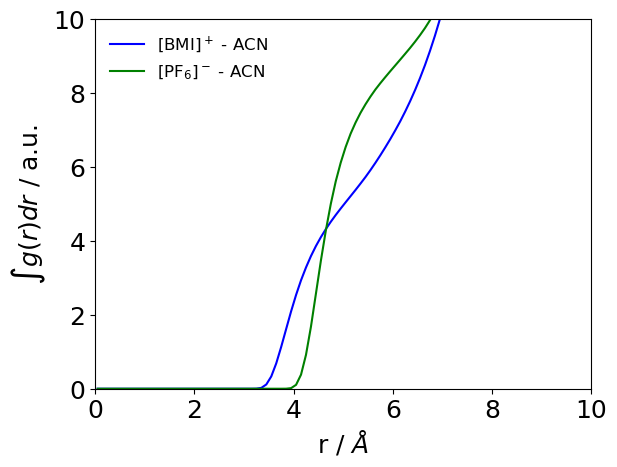}}
  \caption{RDFs of [BMI]$^+$-N\,(ACN) and [PF$_6$]$^-$-Me\,(ACN) (a) and the RDF-integrals (b) for both ions in a 1.5\,M bulk organic electrolyte.\label{S:fig:ele_rdf}}
\end{figure}

\begin{figure}
  \centering
  \subfloat[]{\includegraphics[width=0.5\columnwidth]{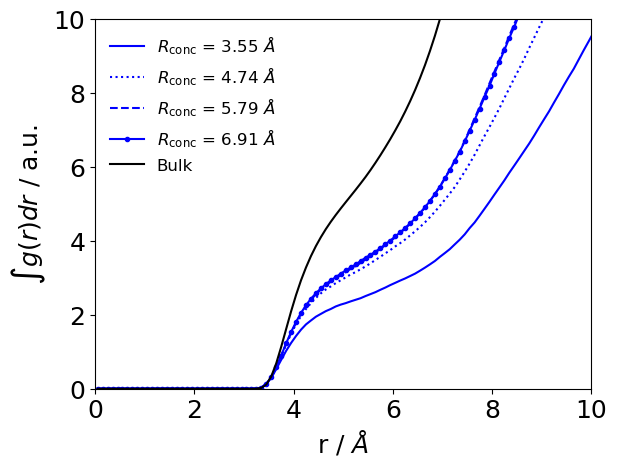}}
  \subfloat[]{\includegraphics[width=0.5\columnwidth]{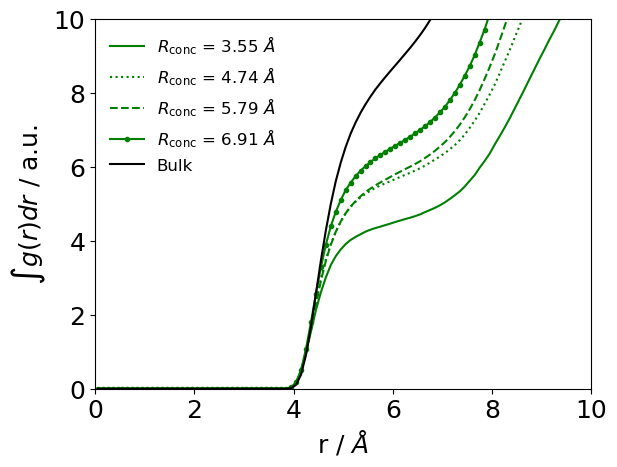}}\\
  \subfloat[]{\includegraphics[width=0.5\columnwidth]{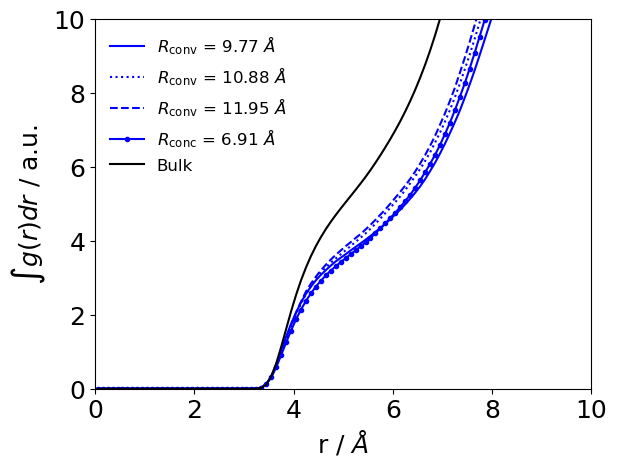}}
  \subfloat[]{\includegraphics[width=0.5\columnwidth]{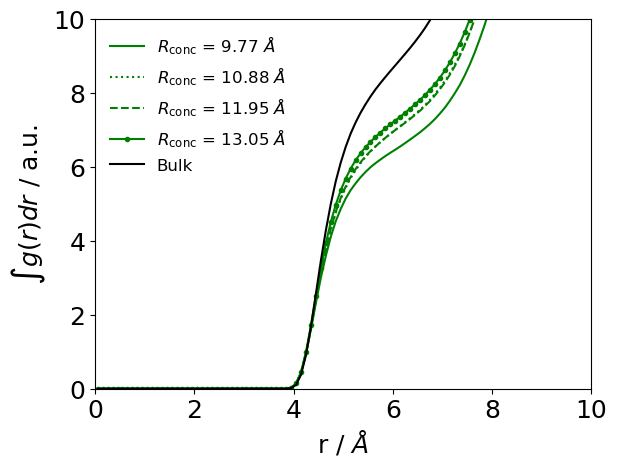}}
  \caption{Integrals of the radial distribution function of [BMI]$^+$-ACN (a,b) and [PF$_6$]$^-$-ACN (c,d) in bulk solution and in the double layer of concave (a,b) and convex (c,d) parts of the electrode with different curvatures, $R_\mathrm{conc}$ and $R_\mathrm{conv}$, respectively. [BMI]$^+$ and [PF$_6$]$^-$ results are obtained from the negative and the positive electrode, respectively. RDFs were obtained from a CONP simulation at 1\,V.} \label{S:fig:solvatshell_curve_1V}
\end{figure}

\begin{figure}
  \centering
  \subfloat[]{\includegraphics[width=0.5\columnwidth]{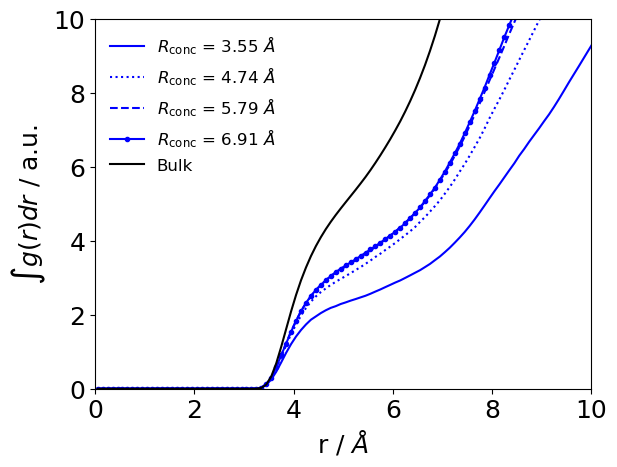}}
  \subfloat[]{\includegraphics[width=0.5\columnwidth]{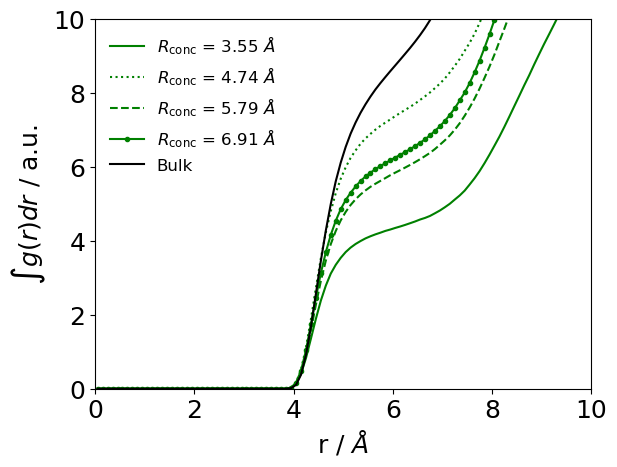}}\\
  \subfloat[]{\includegraphics[width=0.5\columnwidth]{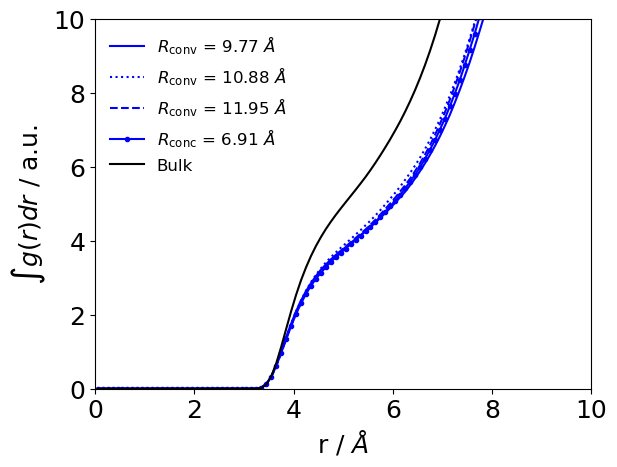}}
  \subfloat[]{\includegraphics[width=0.5\columnwidth]{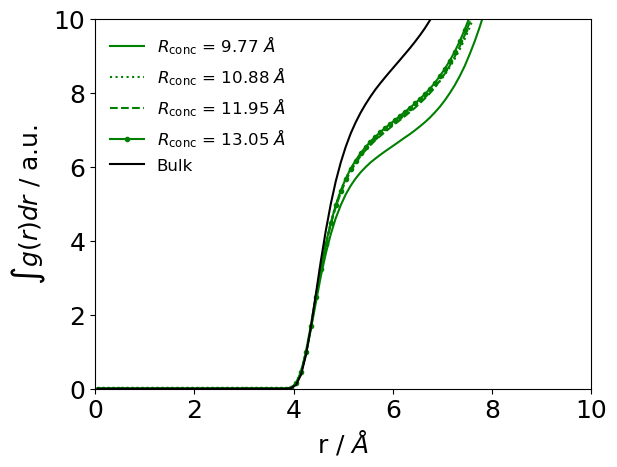}}
  \caption{Integrals of the radial distribution function of [BMI]$^+$-ACN (a,b) and [PF$_6$]$^-$-ACN (c,d) in bulk solution and in the double layer of concave (a,b) and convex (c,d) parts of the electrode with different curvatures, $R_\mathrm{conc}$ and $R_\mathrm{conv}$, respectively. [BMI]$^+$ and [PF$_6$]$^-$ results are obtained from the negative and the positive electrode, respectively. RDFs were obtained from a CONP simulation at 2\,V.} \label{S:fig:solvatshell_curve_2V}
\end{figure}

\begin{figure}
  \centering
  \subfloat[]{\includegraphics[width=0.5\columnwidth]{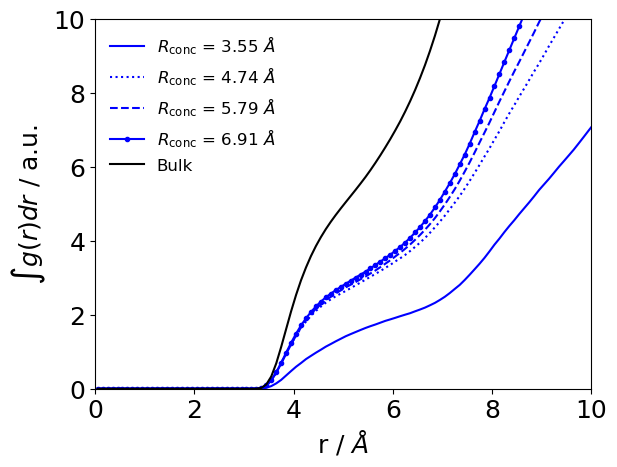}}
  \subfloat[]{\includegraphics[width=0.5\columnwidth]{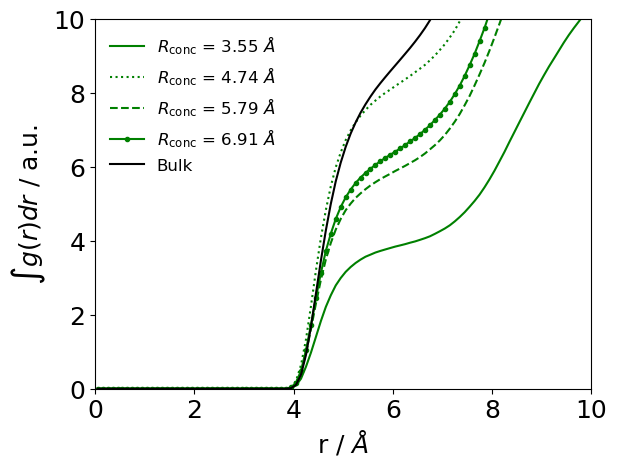}}\\
  \subfloat[]{\includegraphics[width=0.5\columnwidth]{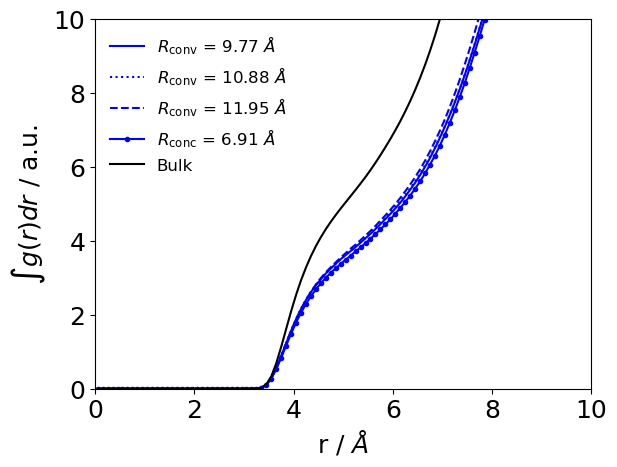}}
  \subfloat[]{\includegraphics[width=0.5\columnwidth]{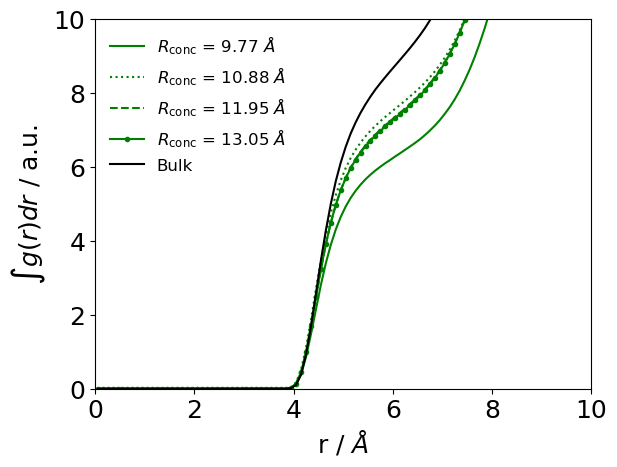}}
  \caption{Integrals of the radial distribution function of [BMI]$^+$-ACN (a,b) and [PF$_6$]$^-$-ACN (c,d) in bulk solution and in the double layer of concave (a,b) and convex (c,d) parts of the electrode with different curvatures, $R_\mathrm{conc}$ and $R_\mathrm{conv}$, respectively. [BMI]$^+$ and [PF$_6$]$^-$ results are obtained from the negative and the positive electrode, respectively. RDFs were obtained from a CONP simulation at 4\,V.} \label{S:fig:solvatshell_curve_4V}
\end{figure}

\begin{figure}
  \centering
  \subfloat[]{\includegraphics[width=0.5\columnwidth]{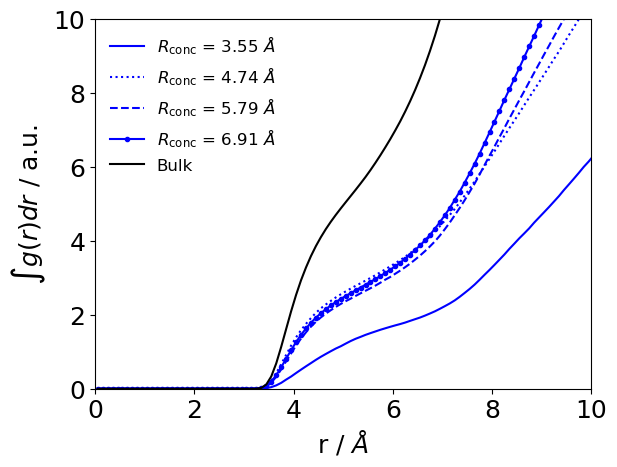}}
  \subfloat[]{\includegraphics[width=0.5\columnwidth]{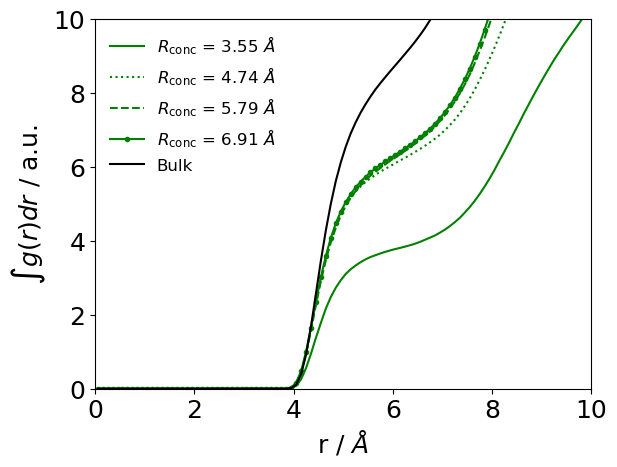}}\\
  \subfloat[]{\includegraphics[width=0.5\columnwidth]{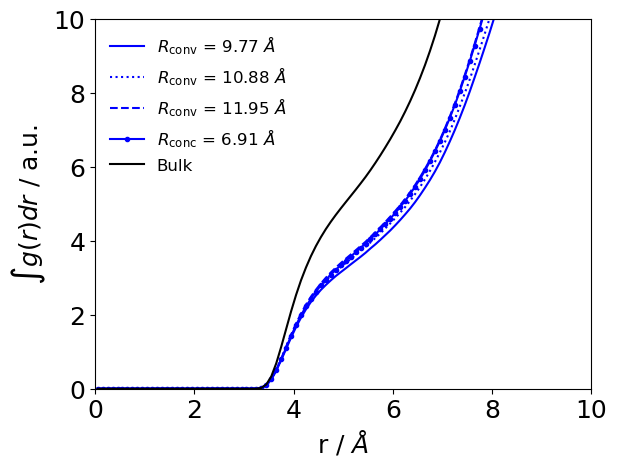}}
  \subfloat[]{\includegraphics[width=0.5\columnwidth]{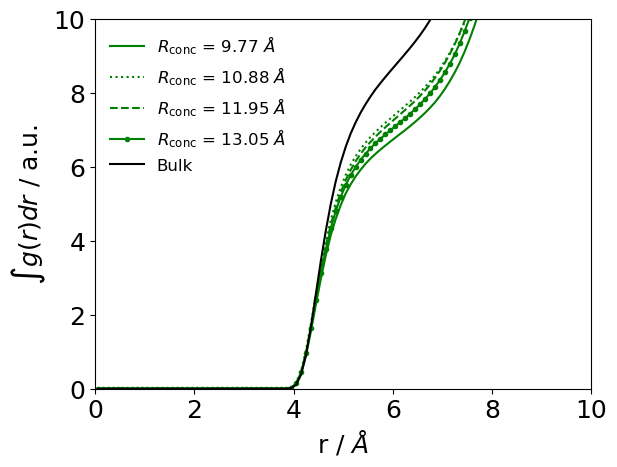}}
  \caption{Integrals of the radial distribution function of [BMI]$^+$-ACN (a,b) and [PF$_6$]$^-$-ACN (c,d) in bulk solution and in the double layer of concave (a,b) and convex (c,d) parts of the electrode with different curvatures, $R_\mathrm{conc}$ and $R_\mathrm{conv}$, respectively. [BMI]$^+$ and [PF$_6$]$^-$ results are obtained from the negative and the positive electrode, respectively. RDFs were obtained from a CONP simulation at 5\,V.} \label{S:fig:solvatshell_curve_5V}
\end{figure}

\FloatBarrier

\subsection{Differential capacitance of flat electrodes}

\begin{figure}
   \centering
   \subfloat[]{\includegraphics[width=0.5\columnwidth]{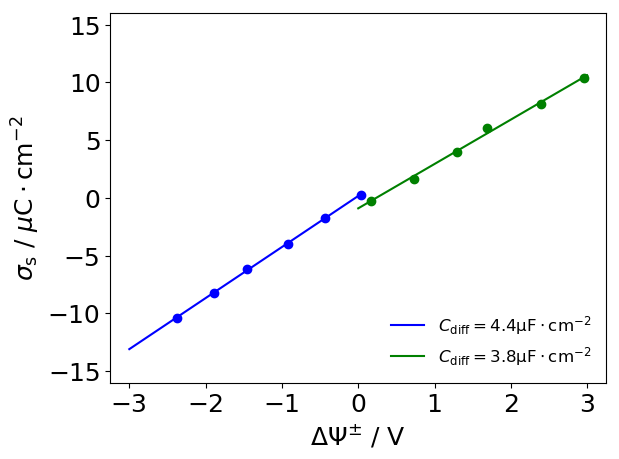}}
   \subfloat[]{\includegraphics[width=0.5\columnwidth]{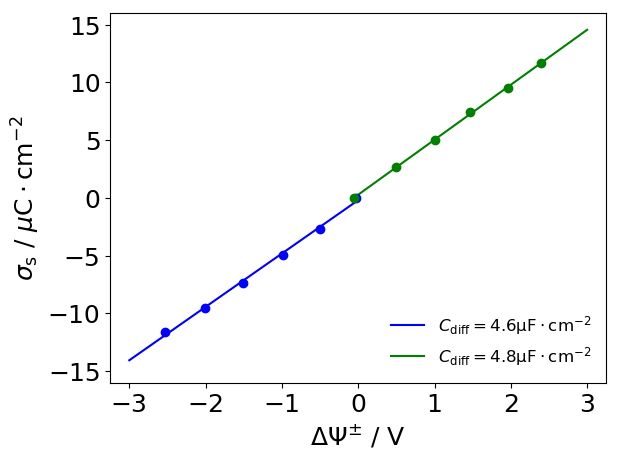}}
   \caption{Estimation of the differential capacitances of flat graphene electrodes for pure [BMI]$^+$[PF$_6$]$^-$ (a) and 1.5\,M [BMI]$^+$[PF$_6$]$^-$ in ACN (b).\label{S:fig:diff_capa_flat}}
\end{figure}

From the literature it is already known, that the use of an organic solvent like ACN leads to an increase of the differential capacitance. Fig.\,\ref{S:fig:diff_capa_flat}(a) shows the differential capacitances of a flat graphitic electrode system with an ionic liquid ([BMI]$^+$ and [PF$_6$]$^-$) and Fig.\,\ref{S:fig:diff_capa_flat}(b) for the same electrode system which is separated by an organic electrolyte (1.5\,M [BMI]$^+$[PF$_6$]$^-$ ACN solution).

\subsection{Number density of different potentials}

The number density was calculated by counting the number of coarse grained components which distance to the electrode carbon atoms is smaller than 4.5\,\AA~(5.0\,\AA). The size of the double layer has been determined as described above.

\begin{figure}
  \centering
  \subfloat[]{\includegraphics[width=0.5\columnwidth]{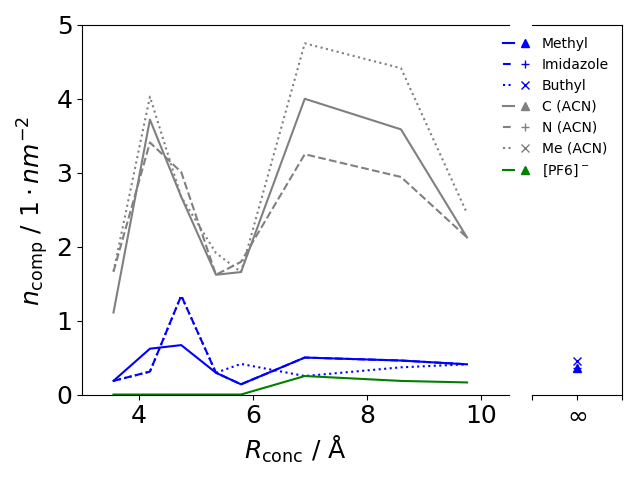}}
  \subfloat[]{\includegraphics[width=0.5\columnwidth]{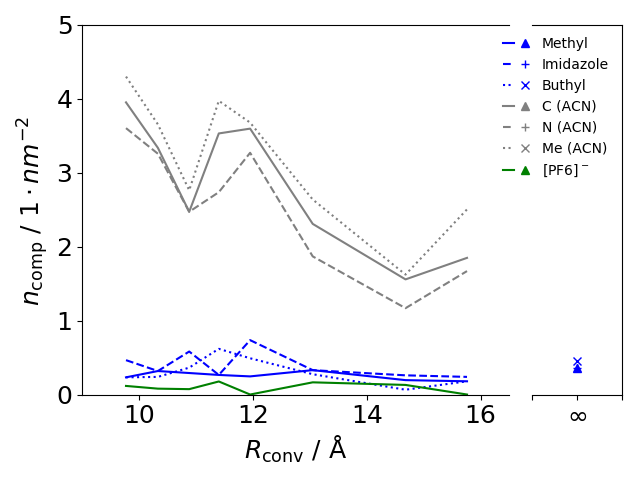}}\\
  \subfloat[]{\includegraphics[width=0.5\columnwidth]{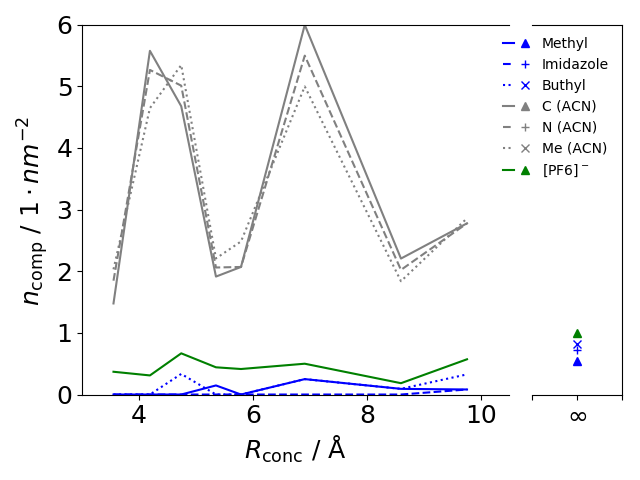}}
  \subfloat[]{\includegraphics[width=0.5\columnwidth]{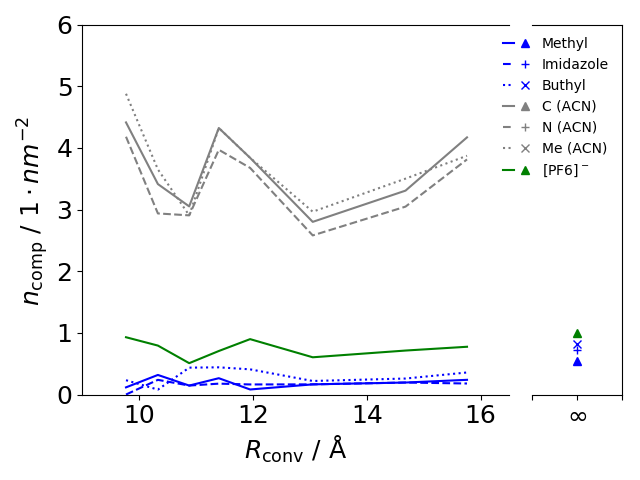}}
  \caption{Number of methyl, imidazole and butyl groups of [BMI]$^+$ [PF$_6$]$^-$ and N, C and methyl group of ACN, $n_\mathrm{comp}$, in the double layer of the negatively charged concave (a) and convex (b) surface area of a negatively charged electrode and for a concave (c) and convex (d) surface area of a positively charged electrode. Results are from a simulation at 1\,V. The size of the double layer was derived from the particle density profile of a supercapacitor with flat graphene electrodes \label{S:fig:number_particle_1V}}
\end{figure}

\begin{figure}
  \centering
  \subfloat[]{\includegraphics[width=0.5\columnwidth]{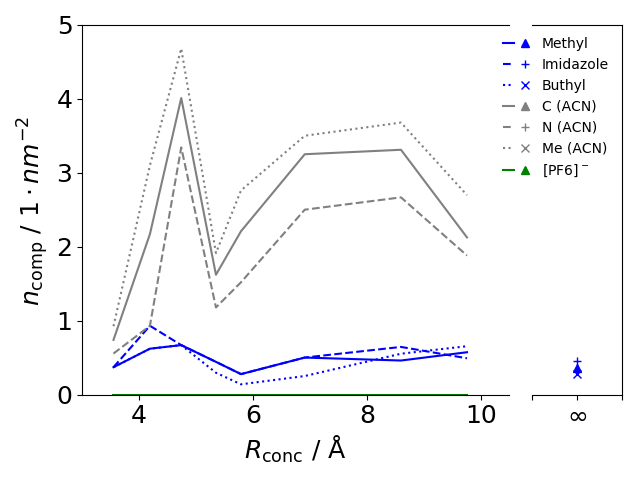}}
  \subfloat[]{\includegraphics[width=0.5\columnwidth]{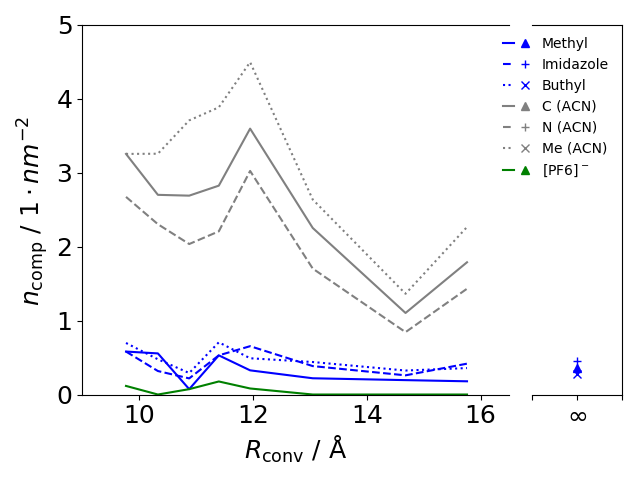}}\\
  \subfloat[]{\includegraphics[width=0.5\columnwidth]{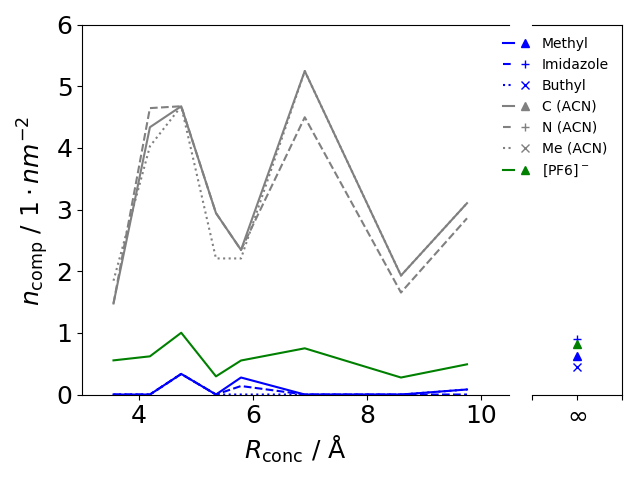}}
  \subfloat[]{\includegraphics[width=0.5\columnwidth]{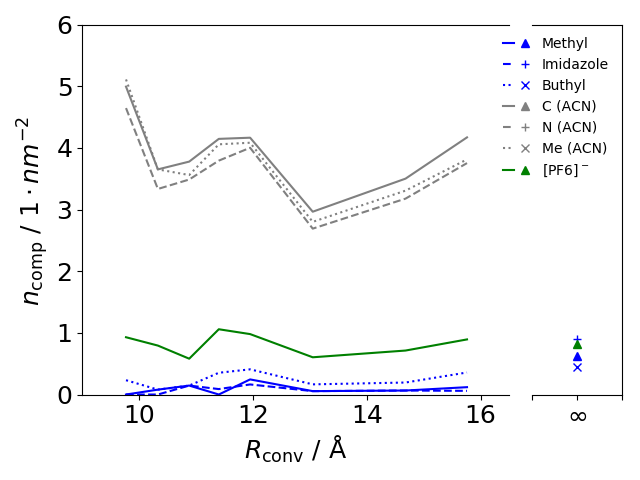}}
  \caption{Number of methyl, imidazole and butyl groups of [BMI]$^+$ [PF$_6$]$^-$ and N, C and methyl group of ACN, $n_\mathrm{comp}$, in the double layer of the negatively charged concave (a) and convex (b) surface area of a negatively charged electrode and for a concave (c) and convex (d) surface area of a positively charged electrode. Results are from a simulation at 2\,V. The size of the double layer was derived from the particle density profile of a supercapacitor with flat graphene electrodes \label{S:fig:number_particle_2V}}
\end{figure}

\begin{figure}
  \centering
  \subfloat[]{\includegraphics[width=0.5\columnwidth]{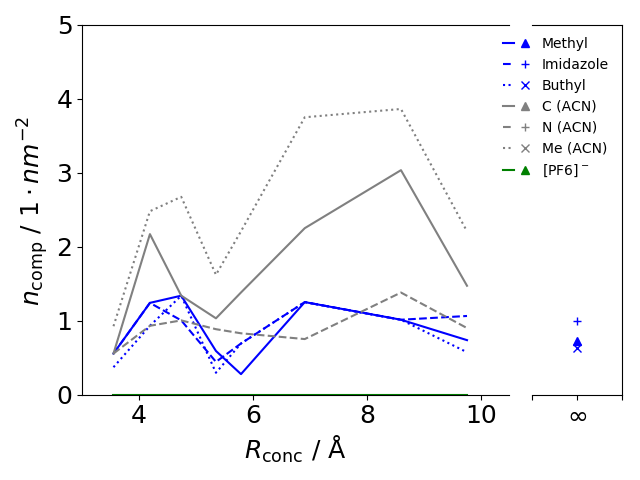}}
  \subfloat[]{\includegraphics[width=0.5\columnwidth]{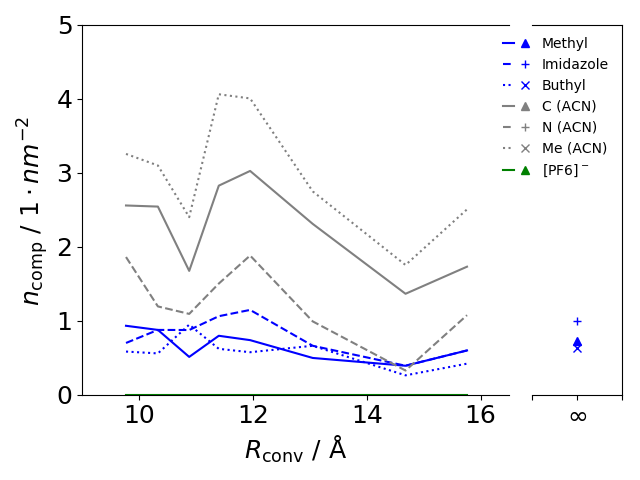}}\\
  \subfloat[]{\includegraphics[width=0.5\columnwidth]{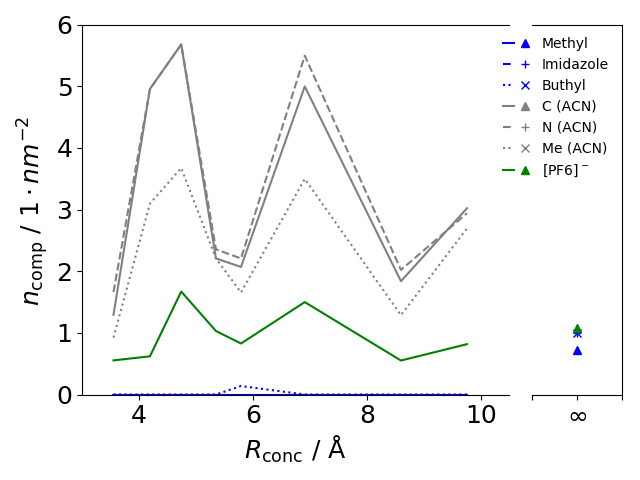}}
  \subfloat[]{\includegraphics[width=0.5\columnwidth]{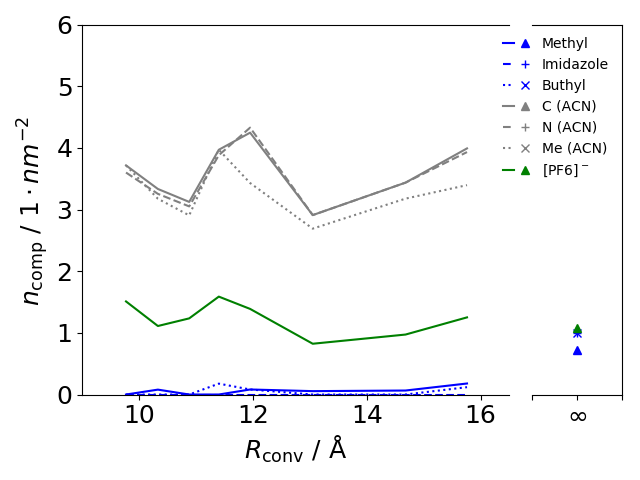}}
  \caption{Number of methyl, imidazole and buthl groups of [BMI]$^+$ [PF$_6$]$^-$ and N, C and methyl group of ACN, $n_\mathrm{comp}$, in the double layer of the negatively charged concave (a) and convex (b) surface area of a negatively charged electrode and for a concave (c) and convex (d) surface area of a positively charged electrode. Results are from a simulation at 4\,V. The size of the double layer was derived from the particle density profile of a supercapacitor with flat graphene electrodes \label{S:fig:number_particle_4V}}
\end{figure}

\begin{figure}
  \centering
  \subfloat[]{\includegraphics[width=0.5\columnwidth]{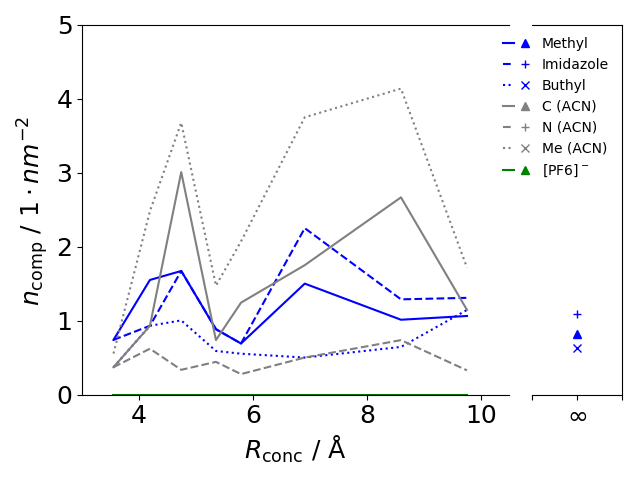}}
  \subfloat[]{\includegraphics[width=0.5\columnwidth]{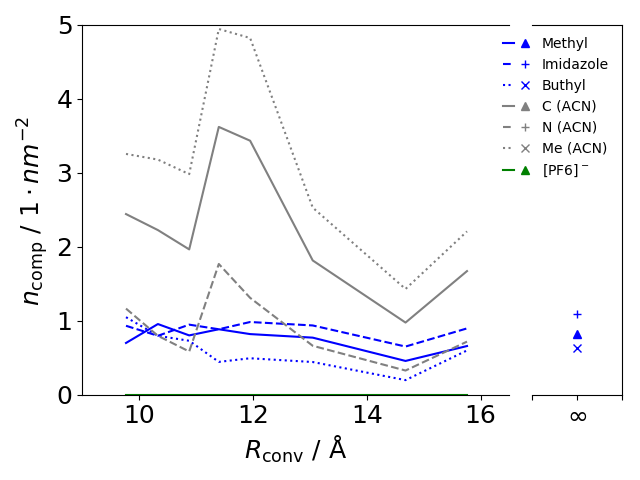}}\\
  \subfloat[]{\includegraphics[width=0.5\columnwidth]{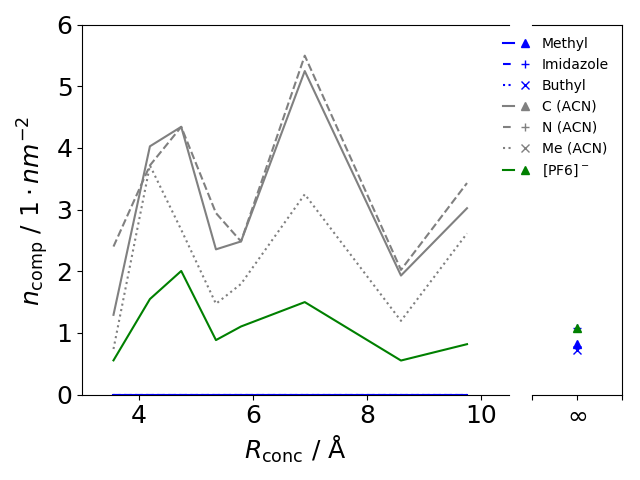}}
  \subfloat[]{\includegraphics[width=0.5\columnwidth]{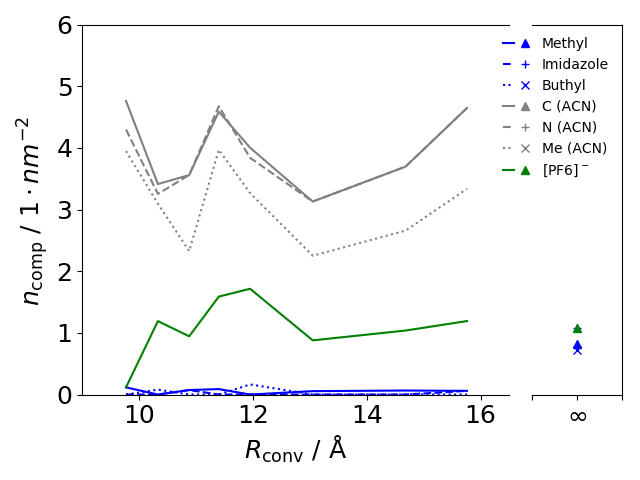}}
  \caption{Number of methyl, imidazole and buthl groups of [BMI]$^+$ [PF$_6$]$^-$ and N, C and methyl group of ACN, $n_\mathrm{comp}$, in the double layer of the negatively charged concave (a) and convex (b) surface area of a negatively charged electrode and for a concave (c) and convex (d) surface area of a positively charged electrode. Results are from a simulation at 5\,V. The size of the double layer was derived from the particle density profile of a supercapacitor with flat graphene electrodes \label{S:fig:number_particle_5V}}
\end{figure}

\newpage

\subsection{Orientation angles of different potentials}

The orientation of the [BMI]$^+$ ions which are adsorbed on the flat as well as on the curved electrode can be defined with three angles:

i) The angle between the line which passes through both the imidazole group and the methyl group and the z-axis of the simulation box. 
ii) The angle between the line which passes through both the imidazole group and the butyl group and the z-axis of the simulation box. 
iii) The angle between the normal of the plane in which the three components of [BMI]$^+$ are located and the z-axis.

\begin{figure}
  \centering
  \subfloat[]{\includegraphics[width=0.3\columnwidth]{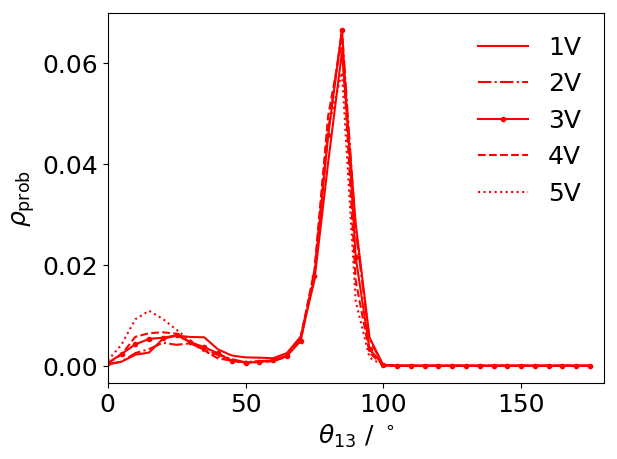}}
  \subfloat[]{\includegraphics[width=0.3\columnwidth]{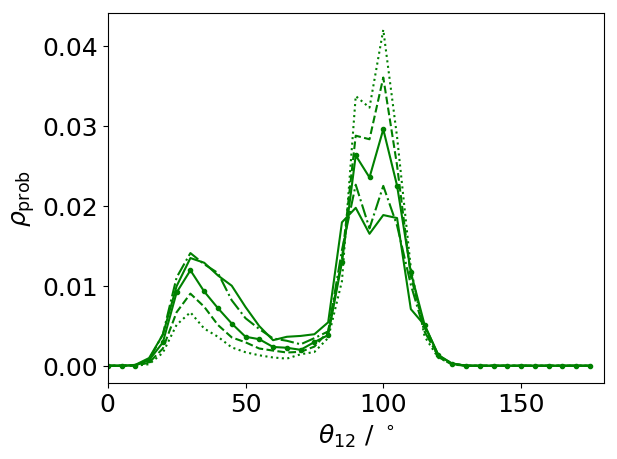}}
  \subfloat[]{\includegraphics[width=0.3\columnwidth]{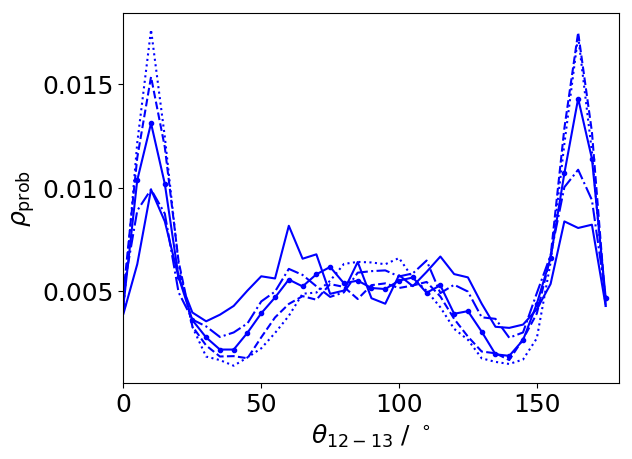}}
  \caption{Histogram of the probability density $\rho_{\mathrm{prob}}$ of the angle between the z axis and the a) line which passes the imidazole and the butyl group of the [BMI]$^+$ $\theta_{13}$ b) line which passes the imidazole and the methyl group $\theta_{12}$, c) normal of the plane in which the three components are located $\theta_{12-13}$ of the [BMI]$^+$ in the double layer of negatively charged flat electrode. The data of different electric potentials are presented. The size of the double layer was derived from RDFs of surface carbon atoms and the imidazole groups.} \label{S:fig:orientation_angle_flat}
\end{figure}

\begin{figure}
  \begin{tabular}{c c}
  \subfloat[]{\includegraphics[width=0.25\columnwidth]{images/angle.png}} &
  \subfloat[]{\includegraphics[width=0.45\columnwidth]{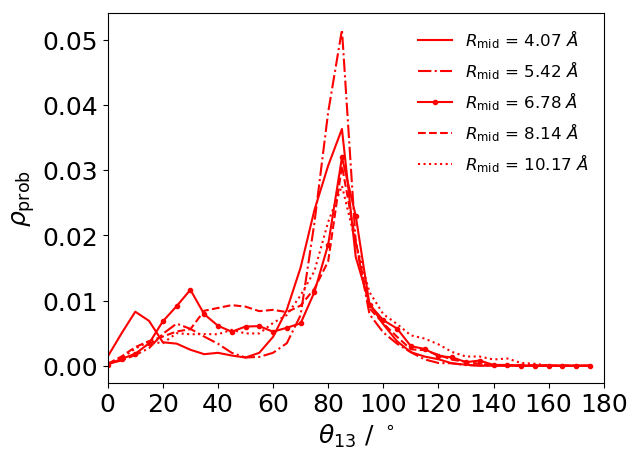}} \\
  \subfloat[]{\includegraphics[width=0.45\columnwidth]{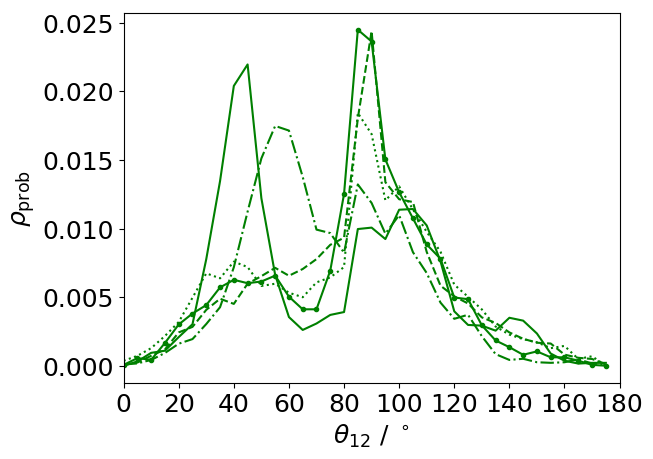}} &
  \subfloat[]{\includegraphics[width=0.45\columnwidth]{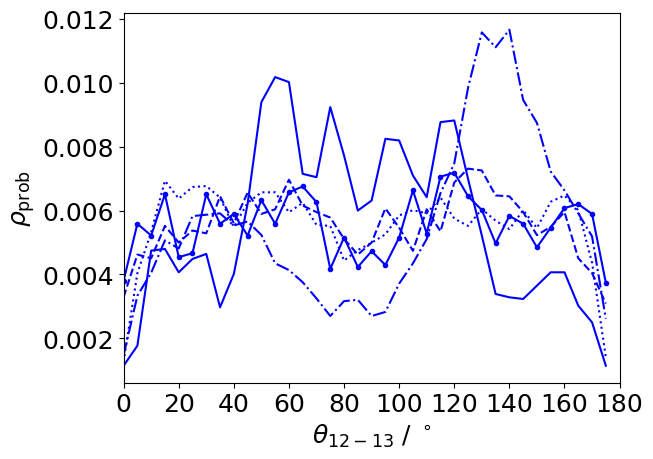}}
  \end{tabular}
  \caption{Histograms of the probability density,$\rho_{\mathrm{prob}}$, of the orientation angle between the z axis and a) the line between the imidazole and the butyl group ($\theta_{12}$) b) the line between the imidazole and the methyl group ($\theta_{13}$) and c) the normal of the plane in which all three groups of the coarsed grained [BMI]$^+$ in the double layer of the negative electrode ($\theta_{12-13}$) are located. Results are from a simulation at 3\,V. The size of the double layer was derived from RDFs between the surface carbon atoms and the imidazole groups. \label{fig:hist_orientation_angle_3V}}
\end{figure}

\begin{figure}
  \centering
  \subfloat[]{\includegraphics[width=0.3\columnwidth]{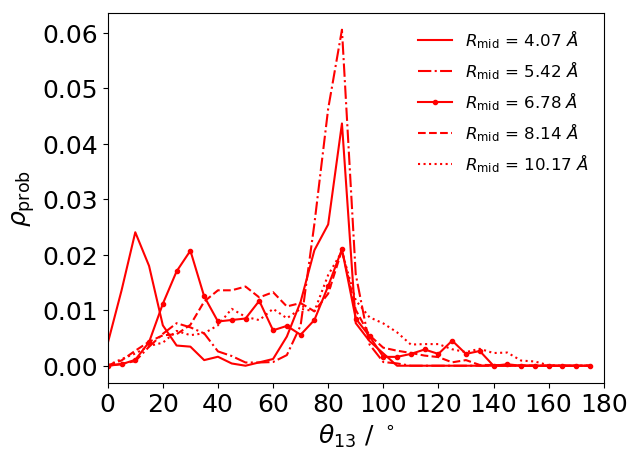}}
  \subfloat[]{\includegraphics[width=0.3\columnwidth]{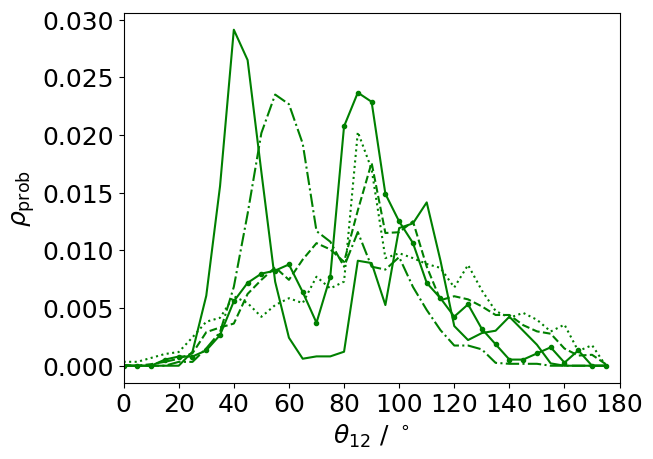}}
  \subfloat[]{\includegraphics[width=0.3\columnwidth]{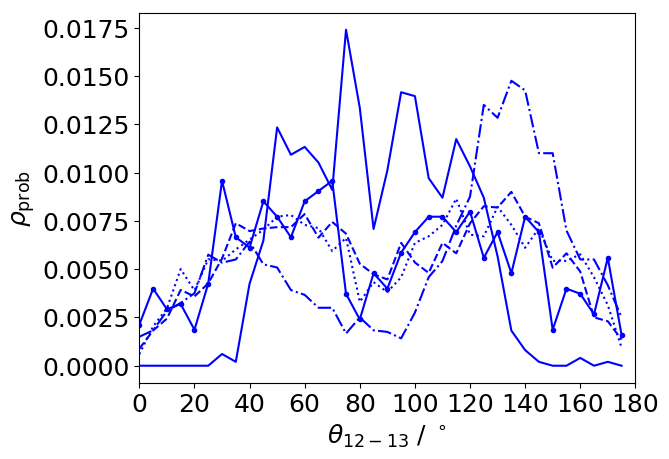}}
  \caption{Histogram of the probability density $\rho_{\mathrm{prob}}$ of the angle between the z axis and the a) line which passes the imidazole and the butyl group of the [BMI]$^+$ $\theta_{13}$ b) line which passes the imidazole and the methyl group $\theta_{12}$, c) normal of the plane in which the three components are located $\theta_{12-13}$ of the [BMI]$^+$ in the double layer of negatively charged concave electrode area at 3\,V. The data of different electric potentials are presented. The size of the double layer was derived from RDFs of surface carbon atoms and the imidazole groups.} \label{S:fig:orientation_angle_3V_conc}
\end{figure}

\begin{figure}
  \centering
  \subfloat[]{\includegraphics[width=0.3\columnwidth]{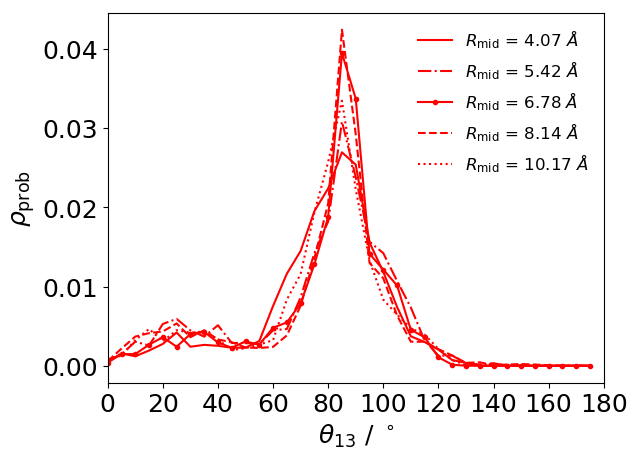}}
  \subfloat[]{\includegraphics[width=0.3\columnwidth]{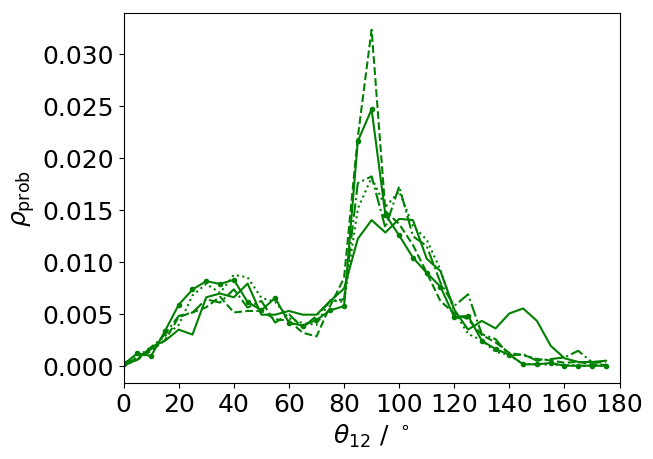}}
  \subfloat[]{\includegraphics[width=0.3\columnwidth]{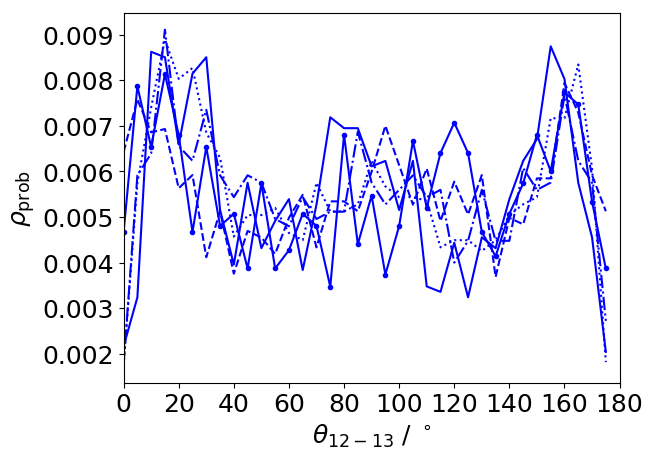}}
  \caption{Histogram of the probability density $\rho_{\mathrm{prob}}$ of the angle between the z axis and the a) line which passes the imidazole and the butyl group of the [BMI]$^+$ $\theta_{13}$ b) line which passes the imidazole and the methyl group $\theta_{12}$, c) normal of the plane in which the three components are located $\theta_{12-13}$ of the [BMI]$^+$ in the double layer of negatively charged convex electrode area at 3\,V. The data of different electric potentials are presented. The size of the double layer was derived from RDFs of surface carbon atoms and the imidazole groups.} \label{S:fig:orientation_angle_3V_conv}
\end{figure}

\newpage
